\documentclass{aa}  
\usepackage{graphicx}
\graphicspath{{./Figures/}} 
\usepackage{xcolor} 
\usepackage{txfonts}
\bibpunct{(}{)}{;}{a}{}{,}
\usepackage{natbib}
\usepackage[colorlinks=true, citecolor=blue]{hyperref}
\usepackage[normalem]{ulem}

\begin{document}

\title{{\scshape SteParSyn}: A Bayesian code to infer stellar atmospheric parameters using spectral synthesis\thanks{Based on observations made with the Mercator Telescope, operated on the island of La Palma by the Flemish Community, at the Spanish Observatorio del Roque de los Muchachos of the Instituto de Astrofísica de Canarias.}\thanks{The full version of Table A.5 is available at the CDS via anonymous ftp
at \url{cdsarc.u-strasbg.fr} (130.79.128.5) or via 
\url{http://cdsarc.u-strasbg.fr/viz-bin/qcat?J/A+A/XXX/YYY}.}}
\subtitle{}
\author{H.~M. Tabernero\inst{1}
        \and
        E. Marfil\inst{2}
        \and 
        D. Montes\inst{2}
        \and 
        J.~I. Gonz{\'a}lez Hern{\'a}ndez\inst{3,4}
 }
\institute{
Centro de Astrobiolog\'\i a (CSIC-INTA), Crta. Ajalvir km 4, E-28850 Torrej\'on de Ardoz, Madrid, Spain\\
\email{htabernero@cab.inta-csic.es}
\and
Departamento de F{\'i}sica de la Tierra y Astrof{\'i}sica \& IPARCOS-UCM (Instituto de F\'{i}sica de Part\'{i}culas y del Cosmos de la UCM), 
Facultad de Ciencias F{\'i}sicas, Universidad Complutense de Madrid, 28040 Madrid, Spain
\and
Instituto de Astrof\'{i}sica de Canarias (IAC), E-38205 La Laguna, Tenerife, Spain
\and
Universidad de La Laguna (ULL), Departamento de Astrof\'{i}sica, E-38206 La Laguna, Tenerife, Spain
}

\date{Received 10 07 2021 / Accepted 10 09 2021}
\abstract{{\scshape SteParSyn} is an automatic code written in Python 3.X designed to infer the stellar atmospheric parameters $T_{\rm eff}$, $\log{g}$, and [Fe/H] of FGKM-type stars following the spectral synthesis method.}
{We present a description of the {\sc SteParSyn} code and test its performance against a sample of late-type stars that were observed with the HERMES spectrograph mounted at the 1.2-m~Mercator Telescope. This sample contains 35 late-type targets with well-known stellar parameters determined independently from spectroscopy. The code is available to the astronomical community in a {\tt GitHub} repository.}
{{\scshape SteParSyn} uses a Markov chain Monte Carlo (MCMC) sampler to explore the parameter space by comparing synthetic model spectra generated on the fly to the observations. The synthetic spectra are generated with an spectral emulator.}
{We computed $T_{\rm eff}$, $\log{g}$, and [Fe/H] for our sample stars and discussed the performance of the code. We calculated an internal scatter for these targets of $-$12~$\pm$~117~K in $T_{\rm eff}$, 0.04~$\pm$~0.14~dex in $\log{g}$, and 0.05~$\pm$~0.09~dex in [Fe/H]. In addition, we find that the $\log{g}$ values obtained with {\sc SteParSyn} are consistent with the trigonometric surface gravities to the 0.1~dex level. Finally, {\sc SteParSyn} can compute stellar parameters that are accurate  down to 50~K, 0.1~dex, and 0.05 dex for T${\rm eff}$, $\log{g}$, and [Fe/H] for stars with $\varv \sin{i}$~$\leq$~30~km~s$^{-1}$.
}
{}

\maketitle
\section{Introduction}

The characterisation of stellar spectra is of great importance to modern astrophysics. It marks the cornerstone for different fields including exoplanets \citep[see][]{valfis05, san13, brew16}, nearby field cosmology \citep[e.g.][]{sil15, bud20}, or even resolved stellar populations in nearby galaxies \citep[see][]{dav17}.  Moreover, the use of automated methods to infer their stellar atmospheric parameters, including effective temperature $T_{\rm eff}$, surface gravity $\log{g}$, and metallicity [Fe/H], allows large surveys to release large databases comprising thousands of stars. Among these surveys are the APO Galactic Evolution Experiment \citep[APOGEE,][]{ahn13}, the GALactic Archaeology with HERMES \citep[GALAH,][]{sil15}, the LAMOST Experiment for Galactic Understanding and Exploration \citep[LEGUE,][]{den12}, the RAdial Velocity Experiment \citep[RAVE,][]{kun17}, the Sloan Extension for Galactic Understanding and Exploration \citep[SEGUE,][]{lee08}, the {\it Gaia}-ESO Survey \citep[GES,][]{gil12},  the WHT Enhanced Area Velocity Explorer  \citep[WEAVE,][]{dal18}, and the 4-metre Multi-Object Spectroscopic Telescope \citep[4MOST,][]{dej19}.\\

In parallel, the census of stars harbouring exoplanets has been steadily increasing over the last decades thanks to the space missions like CoRoT \citep[Convection, Rotation and planetary Transits, ][]{Auv09}, {\it Kepler}  \citep{Koc10, Bor10}, and TESS \citep[Transiting Exoplanet Survey Satellite,][]{ric15}, as well as to the ground-based instruments like the High-Accuracy Radial velocity Planet Searcher \citep[HARPS,][]{may03}, HARPS North \citep[HARPS-N,][]{cos12},  the Calar Alto high-Resolution search for M dwarfs with Exoearths with Near-infrared and optical Echelle Spectrographs \citep[CARMENES,][]{quir20}, and the Echelle Spectrograph for Rocky Exoplanet and Stable Spectroscopic Observations \citep[ESPRESSO,][]{pepe20}. The determination of stellar parameters is also of great interest to exoplanetary science because the masses and radii of the host stars are key to characterising the planets orbiting around them \citep[see, e.g.][]{tor12, san13, brew16, Sou18, Sch19, bru21}. Moreover, they are critical to the high-resolution transmission spectroscopy since they are employed to model the centre-to-limb variation, the limb-darkening, and the Rossiter-McLaughlin effect  \citep[see, e.g.][]{czes15,hoei18,cas20}.\\ 

Broadly speaking, the computation of the stellar atmospheric parameters of FGKM-type stars by means of spectroscopic data can be performed via two different methods: equivalent width (EW) and  spectral synthesis. On the one hand, the EW method uses the strength of several spectral lines to calculate the stellar atmospheric parameters. It employs the standard technique based on the ionisation and excitation balance, taking advantage of the sensitivity of the EWs of \ion{Fe}{i} and \ion{Fe}{ii} lines to the stellar atmospheric parameters \citep[see, e.g.][]{ghe10, tab12, san13}. On the other hand, the spectral synthesis method relies on synthetic spectra used to reproduce the observations using $\chi^2$ fitting algorithms \citep[e.g.][]{valfis05, gar16, tsa18}. The synthetic spectra, all of which might be divided into spectral regions of interest \citep[see e.g.][]{tsa14,brew16}, are finally compared to observations to find the atmospheric model that reproduces the data.\\

These two methods have been extensively reviewed in the literature \citep[see, e.g.][]{allende16b, nis18, jof18, bla19, mar20}. In fact, open-source  implementations of both methods are available to the astronomical community. Regarding the spectral synthesis method, we find the APOGEE Stellar Parameter and Chemical Abundance Pipeline \citep[ASCAP,][]{gar16}, FERRE \citep{FERRE},  MINESweeper \citep{car20}, MyGIsFOS \citep{sbo14}, The Payne \citep{ting19}, and Spectroscopy Made Easy \citep[SME,][]{pis17, val96}, whereas the EW method is implemented in tools such as ARES+MOOG \citep{sou08,san13}, FAMA \citep{mag13}, GALA \citep{muc13}, SPECIES \citep{sot18}, and {\sc StePar} \citep{tab19}. Interestingly enough, other tools such as iSpec \citep{bla14b}, FASMA \citep{and17,tsa20}, and BACCHUS \citep{mas16} are designed to derive the stellar atmospheric parameters using both approaches. Other implementations of the synthetic method rely on Bayesian schemes to calculate the stellar parameters. These schemes represent an improvement over classical $\chi^2$ fitting methods as they can fully explore the probability distribution of the stellar atmospheric parameters associated with an observed spectrum \citep[see, e.g.][]{scho14,starfish}.\\

In this work, we present a description of {\scshape SteParSyn}, written in Python 3.X, which is designed to retrieve stellar atmospheric parameters of late-type stars under the spectral synthesis method. The code was designed to overcome the limitations of the EW method implemented in the {\sc StePar} code \citep{tab19}.
{\sc SteParSyn} represents a step forward towards the analysis of late-type stellar spectra as it relies on a Markov chain Monte Carlo (MCMC) sampler \citep[{\tt emcee}, see][]{emcee} to explore the probability distribution of the stellar atmospheric parameters. In particular, MCMC methods allow us to see the intrinsic parameter degeneracy and provide the uncertainties directly from the sampled probability distribution. In all, the {\sc SteParSyn} code has already been applied to the study and characterisation of late-type stars. In particular, the code has been employed to study stars in open clusters \citep{neg18,alo19,alo20,neg21}, cepheids \citep{loh18}, stars in the Magellanic clouds \citep{tab18}, exoplanet hosts observed under the ESPRESSO GTO programme \citep{bor21,dem21,lil21}, the first super-AGB candidate in our Galaxy \citep[VX~Sgr, see]{tab21}, and stars belonging to the CARMENES GTO sample {\color{blue} (Marfil et al. 2021, submitted).}\\ 
 
This manuscript is divided into five different sections. We describe the {\scshape SteParSyn} code and its internal workflow in Sect.~\ref{sec:description}. In Sect.~\ref{sec:testing}, we test the performance of the code against a sample of late-type stars. We discuss the results and compare them to those obtained in previous works in Sect.~\ref{sec:discussion}. Finally, the conclusions are presented in Sect.~\ref{sec:conclusions}.

\section{Description of the code}
\label{sec:description}

\subsection{Code workflow}
\label{sec:workflow}

{\sc SteParSyn} is a Bayesian code\footnote{The code is available for download at \url{https://github.com/hmtabernero/SteParSyn} under the two-cause BSD license.} designed to sample the posterior probability distribution of stellar parameters associated with an observed spectrum. Prior to the initialisation of the code the user must provide the following input data: (1) an observed spectrum; (2) a grid of synthetic spectra; (3) a list of spectral masks. First, the observed spectrum should be formatted as a plain-text file containing wavelengths, fluxes, and their corresponding uncertainties. Second, the grid should contain the wavelength regions that cover the spectral features under analysis and cover the parameter space relevant to the observations. Third, the list of spectral masks is a compilation of wavelength intervals defined by their centre and width. These masks allow the user to include only atomic features in the analysis. In addition, these spectral masks might be used to avoid parts of the wavelength regions in order to effectively remove 'bad' pixels that can affect the resulting stellar parameters \citep[e.g. cosmic rays and telluric lines; see also][]{valfis05}.\\

\begin{figure}[ht] 
   \centering                       
    \includegraphics[width=0.4\textwidth]{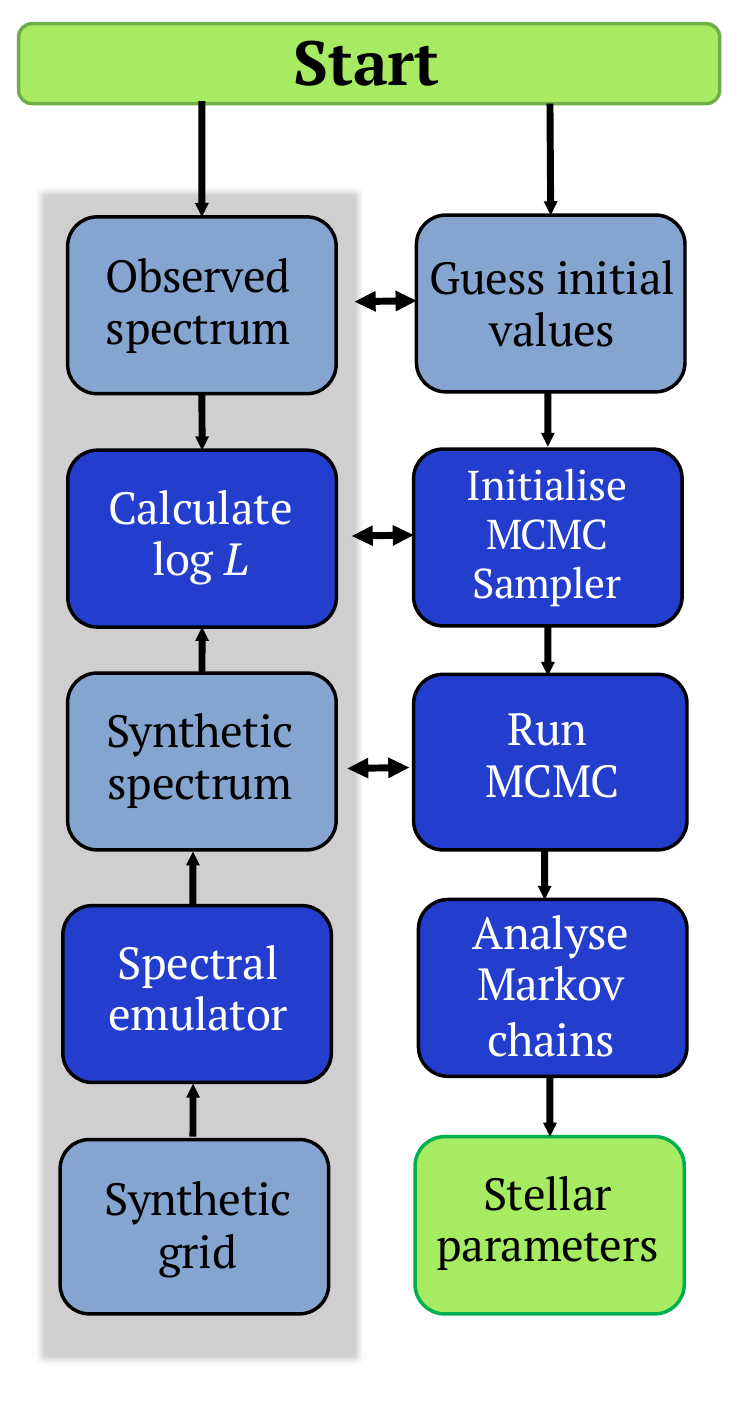}
    \caption{Diagram of the {\sc SteParSyn} internal workflow.}
    \label{workflow}
\end{figure}

After these input data are gathered the code can be initialised following the workflow displayed in Fig.~\ref{workflow}. In the beginning, {\sc SteParSyn} performs a minimisation by means of the {\tt curvefit} subroutine of the {\tt SciPy} Python library \citep{scipy} to obtain a preliminary estimation of the stellar parameters. Then, using these preliminary parameters the code calculates the residuals for each individual wavelength region in order to assess the quality of the user-provided errors. In fact, the observational flux uncertainties calculated from the photon-noise are not enough to explain the difference between models and observations \citep[see, e.g.][]{Hog10,starfish}. These unaccounted-for errors arise from the fact that synthetic spectra do not provide a perfect match to the observations \citep[see, e.g.][]{shet15, tsa18, pass20}. In particular, the precision of the atomic and molecular data employed to generate the model are known to be responsible for these differences. There are millions of features that introduce systematic errors that are perhaps too complex to handle in a one-by-one basis. In spite of this, {\sc SteParSyn} mitigates them by providing a new estimation on the uncertainties of the fluxes. To that aim, the code computes the variance of the residuals from the difference of the synthetic model and the observations for each wavelength region \citep[see ][]{starfish}. These variances are used to compute a robust estimation of the errors for each individual wavelength region.\\

After computing the new robust flux uncertainties, the code creates a set of points centred on the result of the {\tt curvefit} minimisation. These points are evaluated using a likelihood function ($\log{L}$). This function takes into account the synthetic spectrum corresponding to a given point in the parameter space ($F^{\rm syn}_{\lambda}$), the observations ($F^{\rm obs}_{\lambda}$), and their corresponding uncertainties ($\sigma_\lambda$). First, the synthetic spectrum  $F^{\rm syn}_{\lambda}$ is generated by means of an spectral emulator (see Sect.~\ref{emulator}). Second, the synthetic spectrum is multiplied by a scaling factor  given by the median value of the flux points above the third quartile. In other words, the scaling factor allows both the synthetic and observed spectra to be effectively placed at the same flux level. This process is done independently for each wavelength region included in the synthetic grid. Shortly after, the points inside the spectral masks are identified and used to compute the $\log{L}$ according to the following expression:
 
\begin{equation}
\log{L} = \sum_{\lambda} -0.5 \: \left[\left(F^{\rm obs}_\lambda\:-\:F^{\rm syn}_\lambda\right)/\sigma_\lambda\right]^2 -\ln{\left(\sqrt{2\: \pi} \: \sigma_\lambda\right).}  \\
\end{equation}

The aforementioned set of points and their corresponding $\log{L}$ values are used to initialise an MCMC sampler \citep[{\tt emcee},][]{emcee}. The {\tt emcee}\footnote{\url{https://emcee.readthedocs.io/}} package requires the user to provide the number of Markov chains (walkers) to employ and for how long they should be ran (iterations). We recommend to employ 12 walkers for a total of 1\,000 iterations each. Under these conditions, the code takes approximately 40 minutes to complete the sampling (i.e. using a seventh-generation Intel Core i7 processor). Thus, the MCMC sampler has explored the likelihood function that the code translates into the parameters and their uncertainties. At this point, the code delivers the best spectrum that reproduces the features under analysis. Finally, it marginalises the samples drawn from the posterior distribution for each atmospheric parameter and extracts the resulting parameter values and corresponding uncertainties. The former is performed by computing the median and the standard deviation of the marginalised samples. 

\subsection{Synthetic grid}
\label{syngrid}

We provide a grid of synthetic spectra with the public version of {\sc SteParSyn}. We computed this grid thanks to the radiative transfer code {\tt Turbospectrum\footnote{\url{https://github.com/bertrandplez/Turbospectrum2019/}}} \citep{ple12}. In order to generate a synthetic spectrum,{\tt Turbospectrum} needs the following input data: (1) wavelength region to synthesise; (2) a model atmosphere; (3) atomic and molecular data; (4) the chemical composition of the model atmosphere. To that aim, we compiled 296 \ion{Fe}{i} and 31 \ion{Fe}{ii} lines from the following literature sources: \citet{hek07}, \citet{Gen13}, \citet{tsa14}, and \citet{tab19}. Using these iron lines, we produced a list of wavelength regions to synthesise. They were created to cover a 3~{\AA}-wide region centred around each individual line (merging them when necessary, see Table~\ref{linelist}), following the approach taken by \citet{tsa14}. In addition, we chose the MARCS model atmospheres with a standard composition\footnote{\url{https://marcs.astro.uu.se/}} \citep{gus08} spanning from 3\,500 to 7\,000~K in $T_{\rm eff}$, 0.0 to 5.5~dex in $\log{g}$, and $-$2.0 to 1.0~dex in [Fe/H], whereas $\xi$ was set to the values given by the \citet{dutraferr} calibration.\\
 
The atomic line data were taken from the public version of the {\it Gaia}-ESO line list \citep[GES, see][]{hei21}. The GES line list comprises Vienna Atomic Line Database \citep[VALD3,][]{rad15} data plus some highly accurate sources for the atomic parameters on top of hyperfine structure data, all of which were compiled under the framework of the GES collaboration. In addition to the atomic data, we included  data for the following molecular species:  CH, MgH, SiH, CaH, C$_2$, CN, TiO, VO, and ZrO (see Table~\ref{tab:molecules}).  We also removed some transitions from the TiO, VO, and ZrO line lists due to their overwhelming size. Consequently, we followed the automatic recipes provided with the B. Plez molecular databases\footnote{\url{https://nextcloud.lupm.in2p3.fr/s/r8pXijD39YLzw5T}}, which in turn rely on the following expression, as defined by \citet{gra08}:

\begin{equation}
\label{eq:loglik}
\log{({gf_i\cdot\lambda_i})}-\chi_i\cdot\theta > -4 + \max[ \log{({gf_i\cdot\lambda_i})}-\chi_i\cdot\theta]
,\end{equation}

where $\log{gf}_i$ is the oscillator strength of a given transition, $\lambda_i$ is the wavelength in \AA{}, $\chi_l$ is the excitation potential of the lower level in eV, and $\theta = 5\,040/T$ eV$^{-1}$, with $T=3\,500$\,K, and $i$ denotes transition. This approach allowed us to reduce the size of the molecular line lists and use only those transitions that significantly contribute to the spectral synthesis.\\

\begin{table}
\caption{Molecular data employed in this work.}
\label{tab:molecules}
\begin{tabular}{lc}
\hline\hline\noalign{\smallskip}
Molecule & Reference  \\
\hline\noalign{\smallskip}
CH    & \citet{mass14} \\
MgH   & \citet{hin13}  \\
SiH   & \citet{kurSiH}  \\
CaH   &  B. Plez \citep[priv. comm., see][]{hei21}\\
C$_2$ & \citet{bro13}, \citet{ram14} \\
CN    & \citet{sne14}  \\
TiO   & B. Plez \citep[priv. comm., see][]{hei21} \\
VO    & B. Plez \citep[priv. comm., see][]{hei21} \\ 
ZrO   & B. Plez \citep[priv. comm., see][]{hei21}\\
\hline
\end{tabular}
\end{table}

Regarding the chemical composition of the models, we assumed the solar abundances of \citet{asp09} that we scaled according to the composition of each MARCS model atmosphere. In fact, the MARCS models with standard composition take into account the galactic gradient for the so-called $\alpha$-elements (i.e. O, Mg, Si, S, Ar, Ca, and Ti). In consequence, we adopted the following values of the $\alpha$-enhancement (${\rm [\alpha/Fe]}$) as function of metallicity ([Fe/H]): ${\rm [\alpha/Fe]}=0$ if ${\rm [Fe/H]} \ge 0$, ${\rm [\alpha/Fe]}= -0.4{\rm  [Fe/H]}$ if $-1<{\rm [Fe/H]}<0$, and ${\rm [\alpha/Fe]}=0.4$ if ${\rm [Fe/H]}\le -1$.\\

Finally, under the previous assumptions, we ran {\tt Turbospectrum} to compute the synthetic spectra corresponding to the wavelength regions defined in Table~\ref{linelist}. Then, we merged these regions into a single plain text file for each MARCS atmospheric model. These files are later stored into a binary file by means of the {\tt pickle} Python library \citep{python3}. 

\subsection{Spectral emulator}
\label{emulator}

The {\sc SteParSyn} code implements a likelihood function that yields the probability of a synthetic spectrum reproducing the observations (see Eq.~\ref{eq:loglik}). Unfortunately, the stellar atmospheric models are only available at discrete points that do not cover the whole parameter space in a continuous manner. Therefore, we need to be able to produce a synthetic spectrum corresponding to an arbitrary point of the parameter space. The former can be achieved thanks to a spectral emulator, which combines the principal component analysis \citep[PCA, see][]{pca_pearson} decomposition with an interpolation method to reconstruct a synthetic spectrum corresponding to an arbitrary point of the parameter space \citep[see, e.g. ][]{urb08,starfish}. Thus, the PCA allows us to decompose each spectrum as a linear combination according to this formula:

\begin{equation}
\label{recPCA}
F^{\rm syn}_\lambda =  \mu_\lambda + \sigma_\lambda  \sum_{i=1}^{N_{c}} w_i (T_{\rm eff}, \log{g}, {\rm [Fe/H]})\:e^i_\lambda 
.\end{equation}

\begin{figure}[ht]                                                 
   \centering                       
    \includegraphics[width=0.5\textwidth]{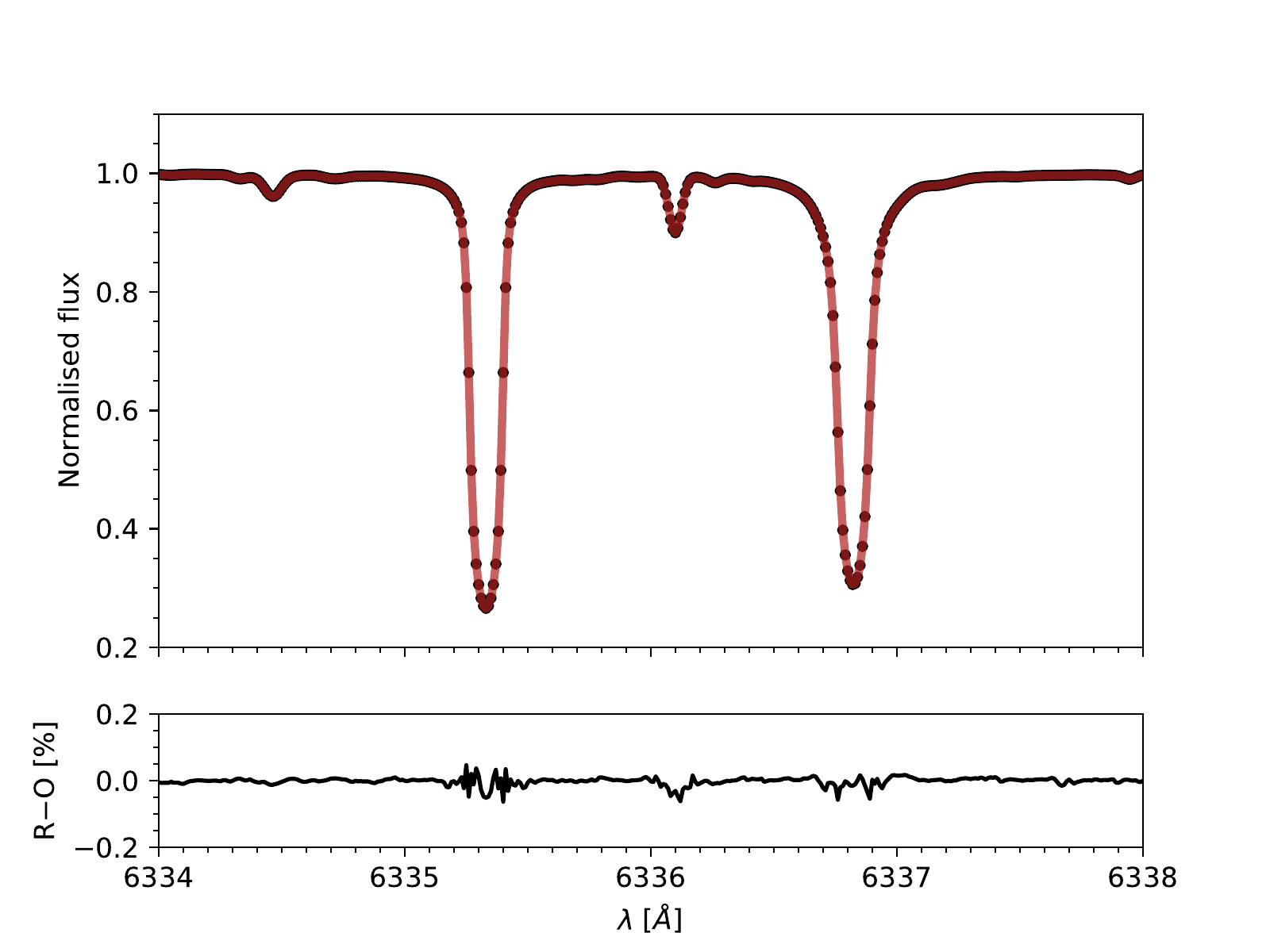}
    \caption{Top panel: Original  synthetic  spectrum (black line) compared to the result of the spectral emulator (shaded red line). Both spectra correspond to  the following stellar parameters: $T_{\rm eff}$~$=$~5750~K, $\log{g}$~4.5~dex, and [Fe/H]~$=$~0.0~dex. Bottom panel: Differences between the reconstructed and original spectra (R-O).}
    \label{fig_recpca}
\end{figure}

Each synthetic spectrum ($F^{\rm syn}_\lambda$) can be expressed as a combination of eigenspectra ($e^i_\lambda$) and weight coefficients ($w_i$) that are given by the PCA. The coefficients $\mu_\lambda$ and $\sigma_\lambda$ represent a linear transformation to the synthetic spectra.  These weights are only known at those points belonging to the grid. Consequently, these weights must be interpolated to produce an spectrum corresponding to an arbitrary point in the parameter space. Then, these interpolated weights can be used to reconstruct a synthetic spectrum corresponding to any point of the parameter space covered by the grid by means of Eq.~\ref{recPCA}. We implemented in the spectral emulator the PCA algorithm of the {\tt scikit-learn} library \citep{sklearn} and the multi-dimensional interpolation algorithm of the {\sc SciPy}  \citep{scipy}. Interestingly enough, the weights obtained using the PCA are not equally important for the reconstruction of each grid spectrum. Interestingly, the grid described in Sect.~\ref{syngrid} was generated using $\approx$~2\,000 MARCS models, and only the top 350 weights were required to bring down the error budget below the 1\% hallmark  (see Fig.~\ref{fig_recpca}).\\

 Finally, the resulting reconstructed synthetic spectrum does not account for the macroturbulence ($\zeta$), the projected rotational velocity ($\varv \sin{i}$), and/or the instrumental resolution. These effects are taken into account by convolving, on the fly, the synthetic spectrum with a broadening kernel. Both $\zeta$ and $\varv \sin{i}$ are too degenerate in the case of FGKM-type stars and they cannot be easily disentangled from each other. One approximation is to model them by means of a single stellar rotation  kernel \citep[e.g.][]{lumba}.  Therefore, {\sc SteParSyn} convolves the synthetic spectrum with a Gray rotation kernel \citep{gra08} to account for both $\zeta$ and $\varv \sin{i}$ followed by a Gaussian kernel with a full width at half maximum (FWHM) corresponding to the resolving power of the observations. 

\section{Code performance}
\label{sec:testing}
\subsection{Test sample}
\label{sample}

 We gathered a sample of 35 late-type stars in order to benchmark {\sc SteParSyn}. Our target list comprises 23 {\it Gaia} benchmark stars \citep{hei15a}, ten late-type stars with measured interferometric angular diameters \citep[see][]{boyafg,boykm}, and two well-known exoplanet host stars \citep[HD~189733 and HD~209458, see][]{boy2pla}. All these targets have accurate $T_{\rm eff}$ and $\log{g}$ values determined independently from spectroscopy. According to their reported parameters, they are spread across the parameter space approximately from 3\,700~K to 6\,700~K in $T_{\rm eff}$, and 0.7~dex to 4.7~dex in $\log{g}$ (see Table~\ref{reference_parameters}). We observed them at the 1.2-m Mercator Telescope\footnote{\url{https://www.mercator.iac.es/}} using the High Efficiency and Resolution Mercator Echelle Spectrograph \citep[HERMES, see][]{hermesmercator} located at the  Observatorio del  Roque de  los  Muchachos  (La Palma, Spain) between 2010 and 2016 (see Table~\ref{obslog}). The HERMES spectrograph covers a wavelength range from 3\,770 to 9\,000~\AA{} with a resolving power of $R$~$=$~$85\,000$ and they were later reduced on-site by means of the HERMES automatic reduction pipeline. We note that these observations include a solar spectrum that was taken by pointing the telescope at the asteroid Vesta \citep[see][]{tab17}.\\

Finally, we shifted these spectra to the laboratory reference frame by correcting for the radial velocity (RV) of the star. We derived the RVs of our targets using the method described in \citep{pepe2002} to  compute  the cross-correlation function (CCF). We sampled the CCF in the range from $-$200 to 200~km~s$^{-1}$ with a step of 0.5~km~s$^{-1}$, using masks that are 1~km~s$^{-1}$ wide and proportionally weighted to their normalised intensity with respect to the  continuum. We assigned each individual weight according to the intensities given by a VALD3 {\tt extract stellar} query with parameters $T_{\rm eff}$~$=$~5\,750~K, $\log{g}$~$=$~4.5~dex, and [Fe/H]~$=$~0.0~dex. We then fitted a Gaussian profile to each CCF to extract the corresponding RVs and calculated their uncertainties by means of the approach described in \citet{zuck03} using the implementation of \citet{bla14b}. We list these RVs and their associated uncertainties in Table~\ref{obslog}.

 \begin{figure}[ht]                             
 \centering                       
    \includegraphics[width=0.4\textwidth]{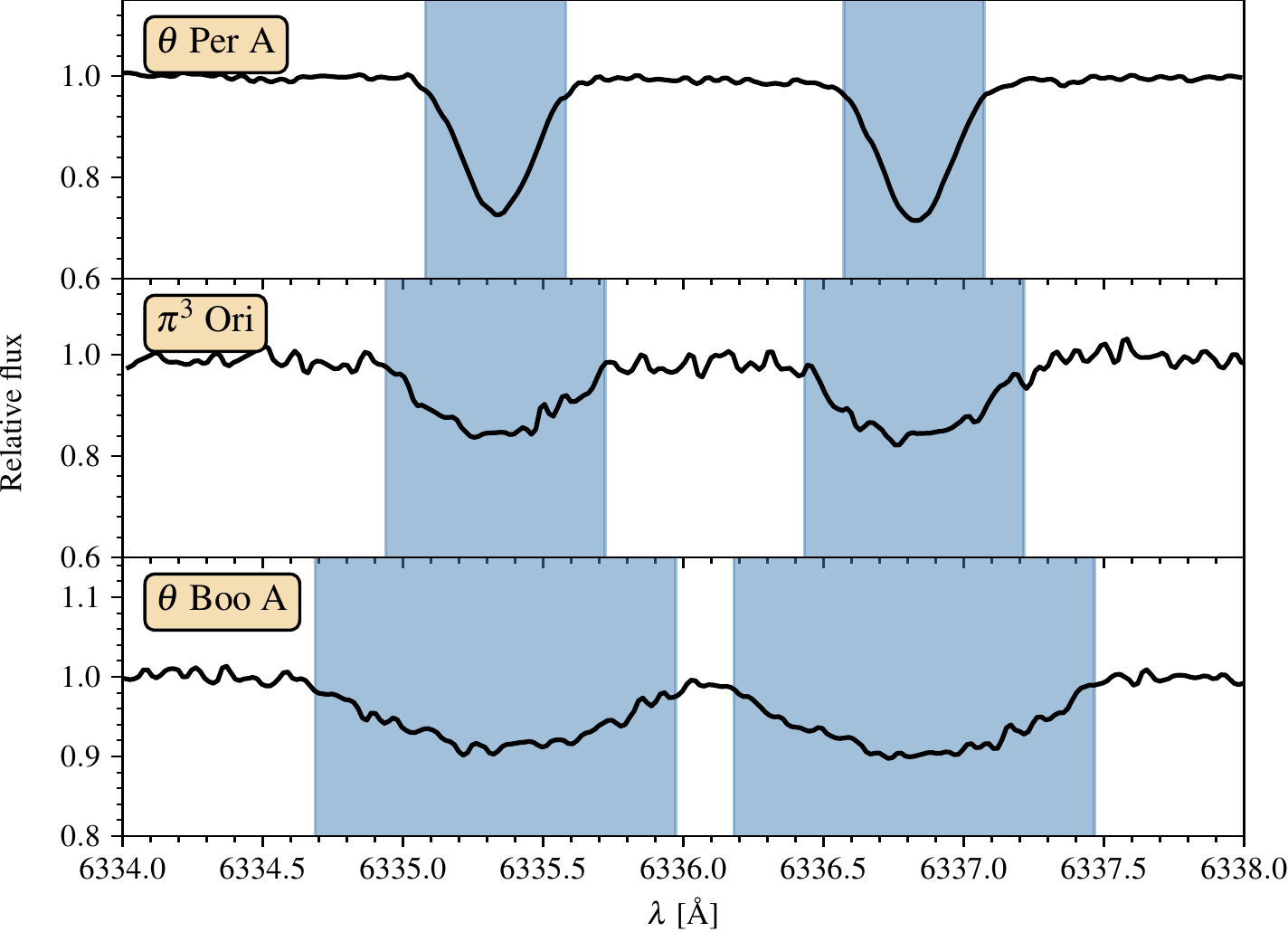}
    \caption{Line masks (shaded blue regions) around the  \ion{Fe}{i} lines at 6335.33 and 6336.82~\AA{} for three F-type stars with different $\varv \sin{i}$ values (see text for details).}
    \label{masks_example}
\end{figure}
 \begin{figure}[ht]                             
 \centering                       
 \includegraphics[width=0.5\textwidth]{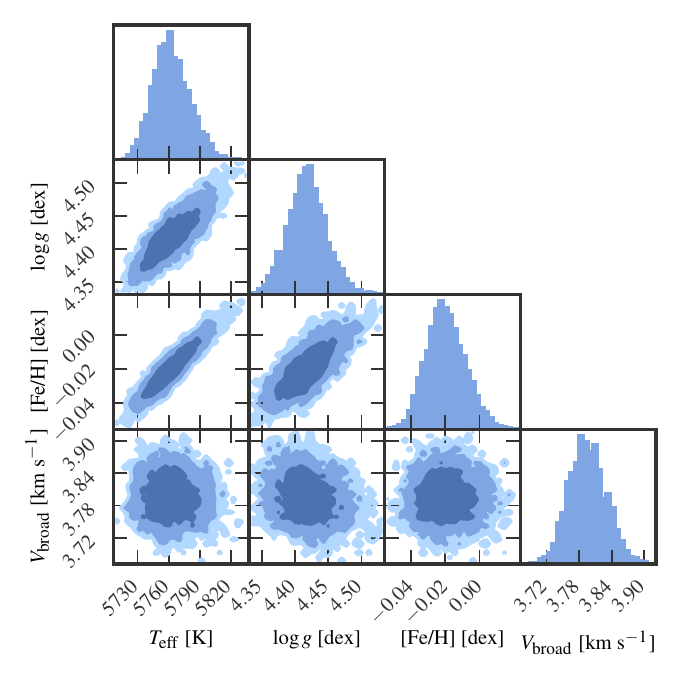}
 \caption{Posterior marginalised distributions corresponding to the following stellar parameters: $T_{\rm eff}$, $\log{g}$,  [Fe/H], and  $V_{\rm broad}$ for the G2~V star 18~Sco. The 1, 2, and 3 $\sigma$ levels are represented by three different colour shades.}.
\label{corner_18sco}
\end{figure}

\begin{figure}[ht]
\includegraphics[width=0.5\textwidth]{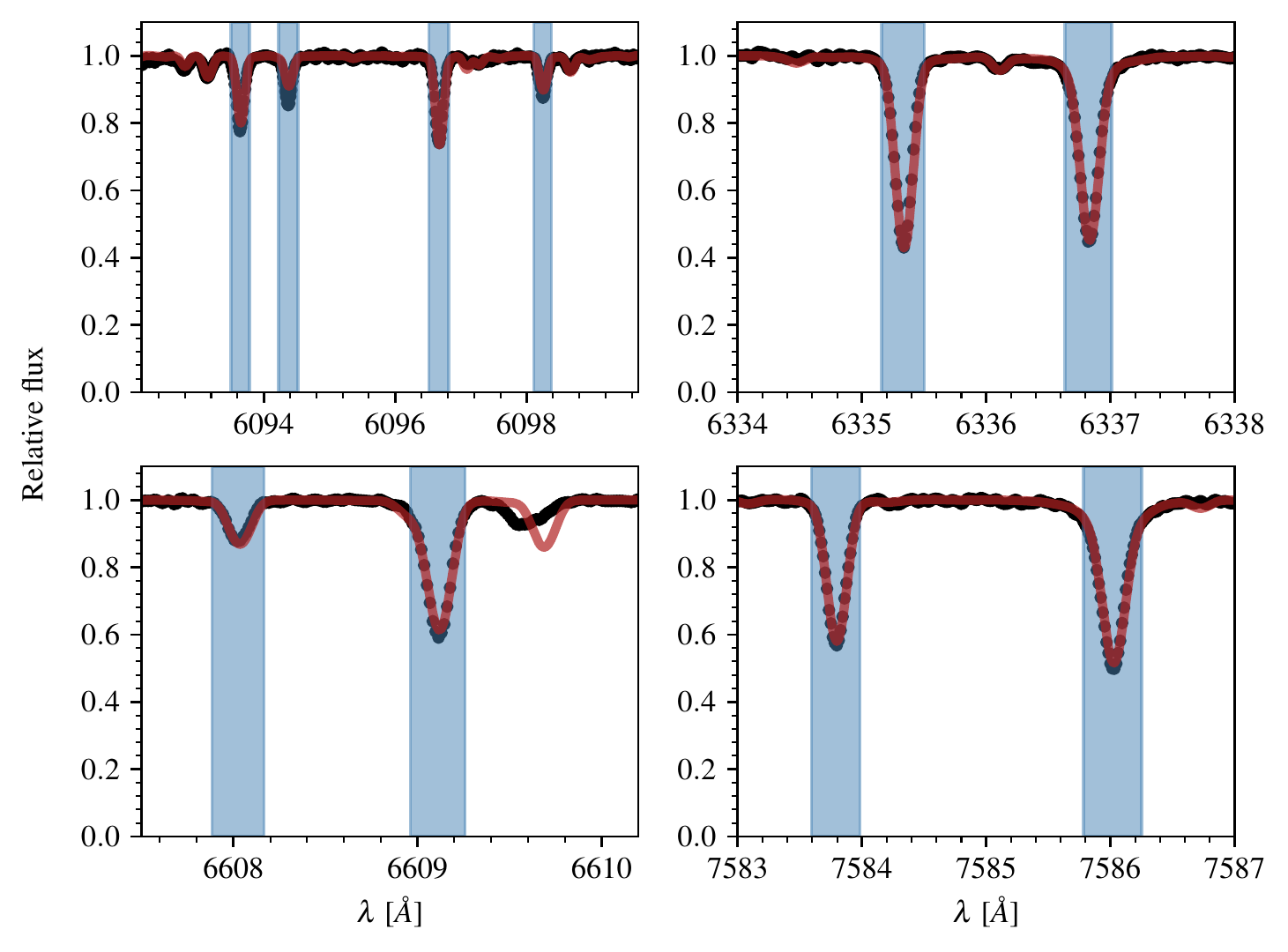}
\caption{Four representative wavelength regions in the spectrum of 18~Sco. The shaded blue areas represent the line masks, whereas the observations are given by the black line, and the best-fitting synthetic spectrum is represented by a red line.}
\label{example_best}
\end{figure}

\subsection{Stellar atmospheric parameters}
\label{parameter}
\begin{figure*}[ht] 
   \centering                       
    \includegraphics[width=\textwidth]{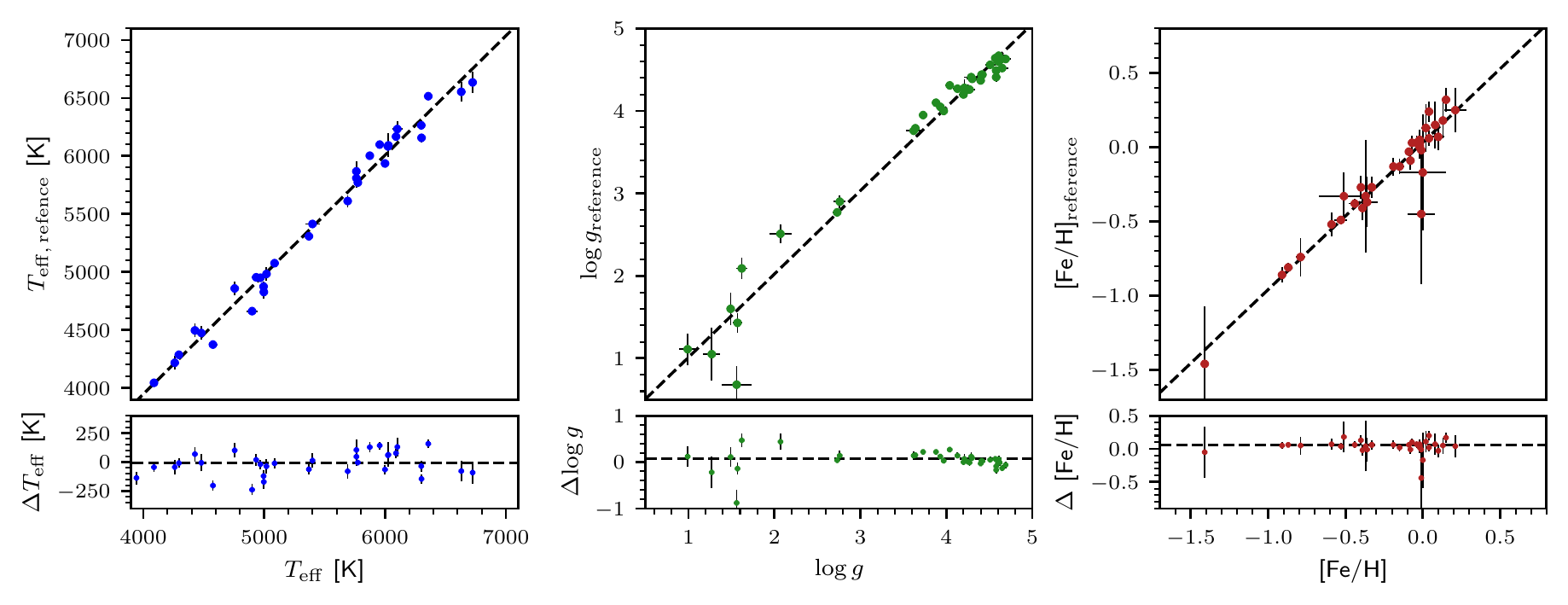}
    \caption{{\sc SteParSyn} results plotted against the reference values given in Table~\ref{reference_parameters}. Upper panels: One-to-one correspondence. Bottom panels: Absolute differences. The dashed black lines in the upper panels correspond to a one-to-one relation, whereas in the bottom panel they correspond to zero.}
    \label{compare}
\end{figure*}

We derived the stellar atmospheric parameters of our targets with {\sc SteParSyn}. Thus, we employed the HERMES observations listed in Table~\ref{obslog}, the synthetic grid described in Sect.~\ref{syngrid}, and a list of spectral masks generated with a custom procedure that we detail below.\\

The list of spectral masks was generated to cover only the \ion{Fe}{i} and \ion{Fe}{ii} lines in the synthetic grid. The underlying idea of deriving the atmospheric parameters using masks around iron lines is to mimic the EW method \citep[see, e.g.][]{sbo14,bla14b}. To build the line masks, we did a Gaussian fit to our iron lines using the Levenberg–Marquardt algorithm (LMA) implemented in the  Python {\tt SciPy} library \citep{scipy}. In summary, our modelling provides the Gaussian parameters of the line under analysis (i.e. continuum level, centre, amplitude, depth, and $\sigma$). The fitting window is 1.5\AA,{} and the continuum level is assumed to be modelled by a constant value. Shortly after, we removed the features whose fitted central wavelengths deviated more than 0.05~\AA{} from their known rest-frame wavelengths. After this filter, the line masks were constructed using an iterative algorithm. We assumed an initial span for the masks of 3$\sigma$ around the centre of each line and adjusted their width to avoid including any neighbouring spectral features. The adjustment of the line masks is performed according to the quality of the Gaussian fit. To that aim, we computed the absolute value of the residuals contained with the initial interval and kept those flux points that deviated no more than 3$\sigma$ from the Gaussian fit. In other words, the initial mask size is effectively reduced until it continuously covers the line of interest and it avoids the contamination of neighbouring features.
However, the Gaussian approximation is no longer valid to reproduce the weak lines when $\varv\sin{i}>15$~km~s$^{-1}$. We note that only $\pi^3$~Ori and $\theta$~Boo~A  have a $\varv \sin{i}$ above the 15~km~s$^{-1}$ mark. In consequence, we followed a different strategy for these two targets to build their line masks. To that aim, we gathered  the lines fitted for the star $\theta$~Per~A and modified them to take into account the stellar rotation. The widths of the new line masks were set to the value given by this formula: $\Delta \lambda$~$=$~$2 \lambda_{\rm o}$~$\varv\sin{i}/c$, where $c$ is the speed of light in km~s$^{-1}$ and $\lambda_{\rm o}$ is the centre of the line given in \AA{}. We illustrate this procedure for two representative \ion{Fe}{i} lines belonging to the spectra of  $\theta$~Per~A, $\pi^3$~Ori, and $\theta$~Boo~A in Fig.~\ref{masks_example}.\\
 
After we computed the line masks corresponding to our selected \ion{Fe}{i,ii} lines, we turned to the broadening parameters of our targets. The {\sc SteParSyn} code models the total line broadening with two components. The first component ($V_{\rm broad}$) accounts for the broadening parameters $\zeta$ and $\varv \sin{i}$. We chose to model both of them with a single rotation kernel following the prescription given by \citet{lumba}. The second component corresponds to the instrumental line spread function (LSF) that we described with a Gaussian function corresponding to the HERMES resolving power (i.e. $R$~$=$~85\,000). The quantity $V_{\rm broad}$ was derived only for those stars with significant broadening parameters. In fact, when $V_{\rm broad}$ is small enough, the total line broadening is entirely dominated by the LSF of the instrument \citep{rei18}. This affects some mid-to-late K dwarf stars in our sample with reported $\varv \sin{i}$ values on or below the noise-driven limit of 2~km~s$^{-1}$ \citep{gra08}. Moreover, they are expected to have small macroturbulent velocities \citep[see][]{brew16} that translate into an LSF-dominated line broadening. Consequently, we fixed the $V_{\rm broad}$ of those K-stars to the $\varv \sin{i}$ reported in the literature (see Table~\ref{reference_parameters}). Then, we ran {\sc SteParSyn} and calculated the stellar atmospheric parameters for our targets. We list the stellar parameters calculated with {\sc SteParSyn} in Table~\ref{steparsyn_results}. We also represent the marginalised results provided by {\sc SteParSyn} for the G2~V star 18~Sco in Fig.~\ref{corner_18sco} and the best fit to some representative iron lines in Fig.~\ref{example_best}.\\

\begin{figure}[ht]                                                  
   \centering                       
    \includegraphics[width=0.5\textwidth]{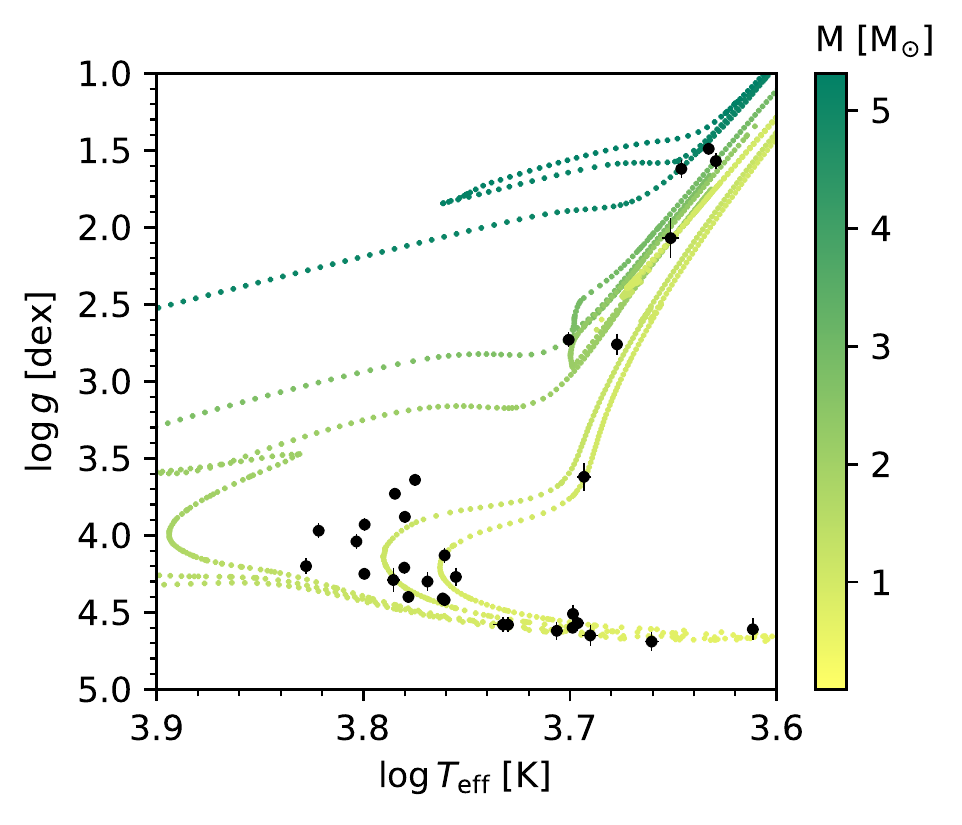}
    \caption{Kiel diagram ($\log{g}$ vs. $\log{T_{\rm eff}}$) for all the spectra analysed alongside the PARSEC isochrones for 0.1, 0.5, 1, 5, and 10 Ga for solar metallicity \citep{bre12}.}
    \label{kiel_syn}
    \end{figure}

\begin{figure}[ht]                                      
   \centering                       
    \includegraphics[width=0.5\textwidth]{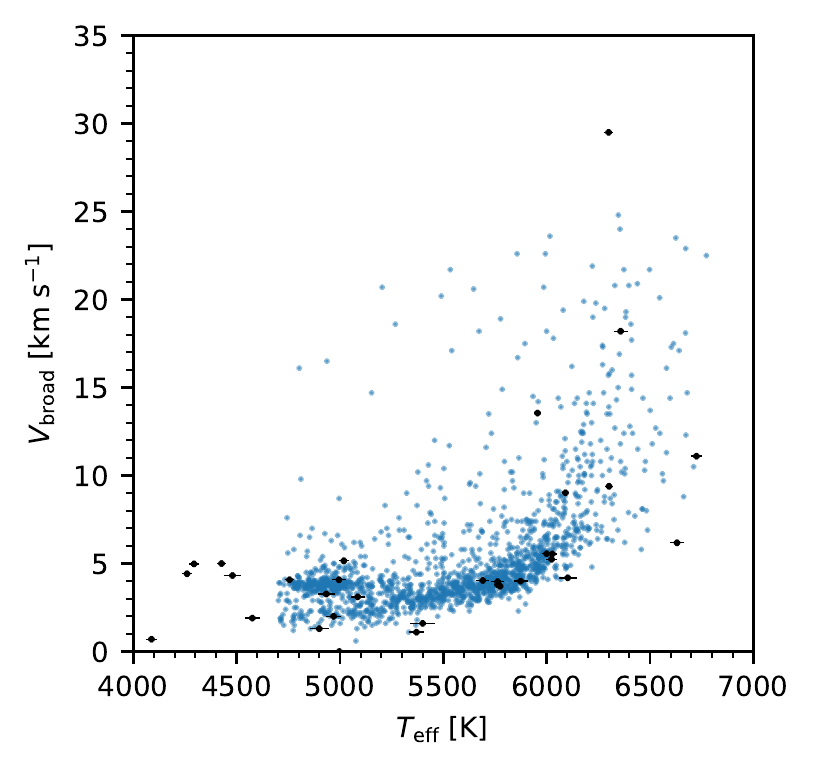}
    \caption{Comparison of the  $V_{\rm broad}$ derived in this work represented by black circles and the values derived by \citet{brew16} are displayed with blue circles.}
    \label{vsini}
\end{figure}

We then compared our results to the reference parameters (see Table~\ref{reference_parameters}) and represented them in Fig.~\ref{compare}. Then, following \citet{tab19}, we performed 10\,000 Monte Carlo simulations on our data  in the hope of assessing possible sources of tentatively systematic offsets. We found the following differences with the reference parameters: $-$12~$\pm$~117~K in $T_{\rm eff}$, 0.04~$\pm$~0.14 dex in $\log{g}$, and 0.05~$\pm$~0.09 dex in [Fe/H]. These differences show that SteParSyn can reproduce the reference values for our sample stars since they are compatible with zero in all instances. In parallel, we populated a Kiel diagram (see Fig.~\ref{kiel_syn}) with our $T_{\rm eff}$ and $\log{g}$ values alongside the PARSEC isochrones \citep{bre12}. The isochrones we represent in the Kiel diagram encompass our points and they are well-behaved even for the coolest stars in the main-sequence, contrary to what has been found in previous works \citep[see ][]{mor14,mon18,tsa19,bru21}. Finally,  we compared our $V_{\rm broad}$ values to those of other late-type stars. Thus, we gathered the broadening values calculated by \citet{brew16} and represented them against our own as a function of $T_{\rm eff}$ in Fig.~\ref{vsini}. From this comparison, we see that our calculated $V_{\rm broad}$ values follow the behaviour of the literature values for other late-type stars reported in \citep{brew16}. 

\subsection{Trigonometric gravities, radii, and masses}
\label{paragrav}

\begin{figure}[ht]                                      
   \centering                       
    \includegraphics[scale=1.0]{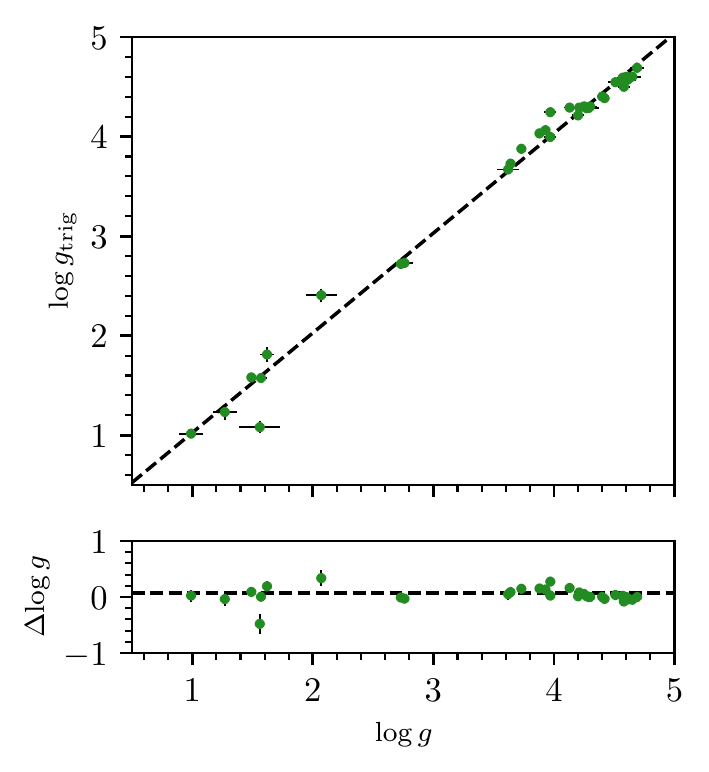}
    \caption{ {\sc SteParSyn} -based surface gravities plotted against the trigonometric values based on the parallaxes listed in Table~\ref{steparsyn_results}.}
    \label{compare_grav}
\end{figure}

The surface gravities derived with spectroscopic data are sometimes not entirely in agreement with those calculated using evolutionary models, visual magnitudes, and parallaxes \citep{tsa19,bru21}. The use of evolutionary models provides us with an alternative method that can be used to assess the precision of the spectroscopic $\log{g}$ values. For these reasons, we used the PARAM web interface\footnote{\url{http://stev.oapd.inaf.it/cgi-bin/param}} \citep{sil06} alongside the $T_{\rm eff}$ and [Fe/H] inferred with {\sc SteParSyn}, the {\it Gaia} EDR3 \citep{EDR3} or {\it Hipparcos}  \citep{newhip} parallax and their visual magnitude ($V$), and the PARSEC stellar evolutionary tracks and isochrones \citep{bre12}. From this information, we calculated the so-called trigonometric surface gravity ($\log{g_{\rm trig}}$), mass ($M_{*}$), and radius ($R_{*}$) for our targets (see Table~\ref{steparsyn_results}). We compare the spectroscopic gravities derived with {\sc SteParSyn} against their trigonometric counterparts in Fig.~\ref{compare_grav}. Again, we ran 10\,000 MC simulations  and computed a  mean difference of -0.03~$\pm$~0.11~dex, which is in turn compatible with zero at the 1$\sigma$ level. This difference implies that our spectroscopic surface gravities are fully consistent  with the PARAM trigonometric values at the 0.1~dex mark.

\subsection{{\sc SteParSyn} vs. {\sc StePar}}
\label{sec:SYNvsEW}

The analysis described in this work uses a selection of \ion{Fe}{i} and \ion{Fe}{ii} lines to derive the stellar parameters of our sample stars. In ideal terms, the synthetic method should deliver the same stellar parameters as the EW method. To quantify the former statement, we calculated the stellar parameters for those targets that can be analysed under the EW method with the {\sc StePar} code\footnote{\url{https://github.com/hmtabernero/StePar/}}. To that aim, we selected the stars of our sample with $\varv \sin{i}$~$<$~15~km~s$^{-1}$ and spectral types between F6 and K4 \citep[see ][]{tab19}. Using these criteria, we finished with a sub-sample of 28 stars. Shortly after, we used the  {\tt ARES}\footnote{\url{https://github.com/sousasag/ARES}} \citep[version 2, see][]{sou15} code to measure the EW of the Fe lines listed  listed in Table~\ref{linelist} for these stars. We calculated the stellar parameters for this sub-sample with {\sc StePar} (see Table~\ref{stepar_results}). We then compared the results of {\sc SteParSyn} and {\sc StePar}. Thus, we retrieve the following differences between them: 40~$\pm$~87~K for $T_{\rm eff}$, $-$0.04~$\pm$~0.21~dex for $\log{g}$, and 0.04~$\pm$~0.08~dex for [Fe/H]. These differences are compatible within the error bars. Therefore, {\sc StePar} and {\sc SteParSyn} deliver compatible results when we analyse the same observational data. Regarding the uncertainties on the parameters, {\sc StePar} gives the following average values: $\Delta$~$T_{\rm eff}$~$=$~50~K, $\Delta$~$\log{g}$~$=$~0.13~dex and $\Delta$~[Fe/H]~$=$~0.04~dex; whereas, for {\sc SteParSyn} we find 29~K, 0.06~dex, and 0.03~dex. Thus, the {\sc SteParSyn} uncertainties are smaller by a factor 2  for $T_{\rm eff}$ and $\log{g}$, whereas both codes provide similar errors on [Fe/H].

\section{Discussion}
\label{sec:discussion}

The stellar atmospheric parameters obtained with {\sc SteParSyn} for our target stars are reliable down to the 1$\sigma$ level with the reference parameters (Table~\ref{reference_parameters}), the trigonometric gravities of Table~\ref{steparsyn_results}, and the parameters provided by the EW method implemented in {\sc StePar} (see Table~\ref{stepar_results}). Roughly speaking, these differences are at the level of 100~K for $T_{\rm eff}$, 0.1~dex for $\log{g}$, and 0.1~dex for [Fe/H] (see Table~\ref{mc_table}). However, these average differences alone are not enough to explore hidden systematic uncertainties tied to the  {\sc SteParSyn} stellar parameters. Thus, we calculated the correlation between the individual differences against the parameters themselves following the approach described in \citet{tab19}. We performed 10\,000 Monte Carlo (MC) simulations on our data to compute both Pearson and Spearman correlation coefficients ($r_p$ and $r_s$, respectively) and their uncertainties. We provide the results of these MC simulations in Table~\ref{mc_table}. These simulations reveal that the correlation between differences and a given parameter are small. Consequently, the average differences found for the stellar atmospheric only correspond to systematic uncertainties intrinsic to  the {\sc SteParSyn} code.\\

We notice that {\sc SteParSyn} [Fe/H] values for both $\alpha$~Cet and $\gamma$~Sge are not entirely in agreement with reference parameters. However, their reference [Fe/H] values largely have error bars at the level of 0.4--0.5~dex, and they are not well-constrained. A similar argument applies to the surface gravities of some giant stars. Moreover, reproducing the surface gravities of the giant stars is by no means an easy task. Stars such as $\mu$~Leo, $\alpha$~Cet, and HD~107328 have surface gravities that are uncertain for a number of currently known reasons. In fact, according to \citet{hei15b} their masses have large uncertainties ranging from approximately 0.5 to 1.0 $M_\odot$ that translate into surface gravities that are not well-defined. In spite of those uncertain $\log{g}$ values, {\sc SteParSyn} manages to produce reliable parameters for our target stars. In addition, their $T_{\rm eff}$ and $\log{g}$ values are in agreement with evolutionary models in the Kiel diagram (see Fig.~\ref{kiel_syn}). These findings are  reinforced by the fact that {\sc SteParSyn} produces values of $V_{\rm broad}$ consistent with the behaviour of those calculated by \citet{brew16}, as we show in Fig.\ref{vsini}.\\
 
\begin{figure}[ht]                                      
   \centering                       
    \includegraphics[width=0.45\textwidth]{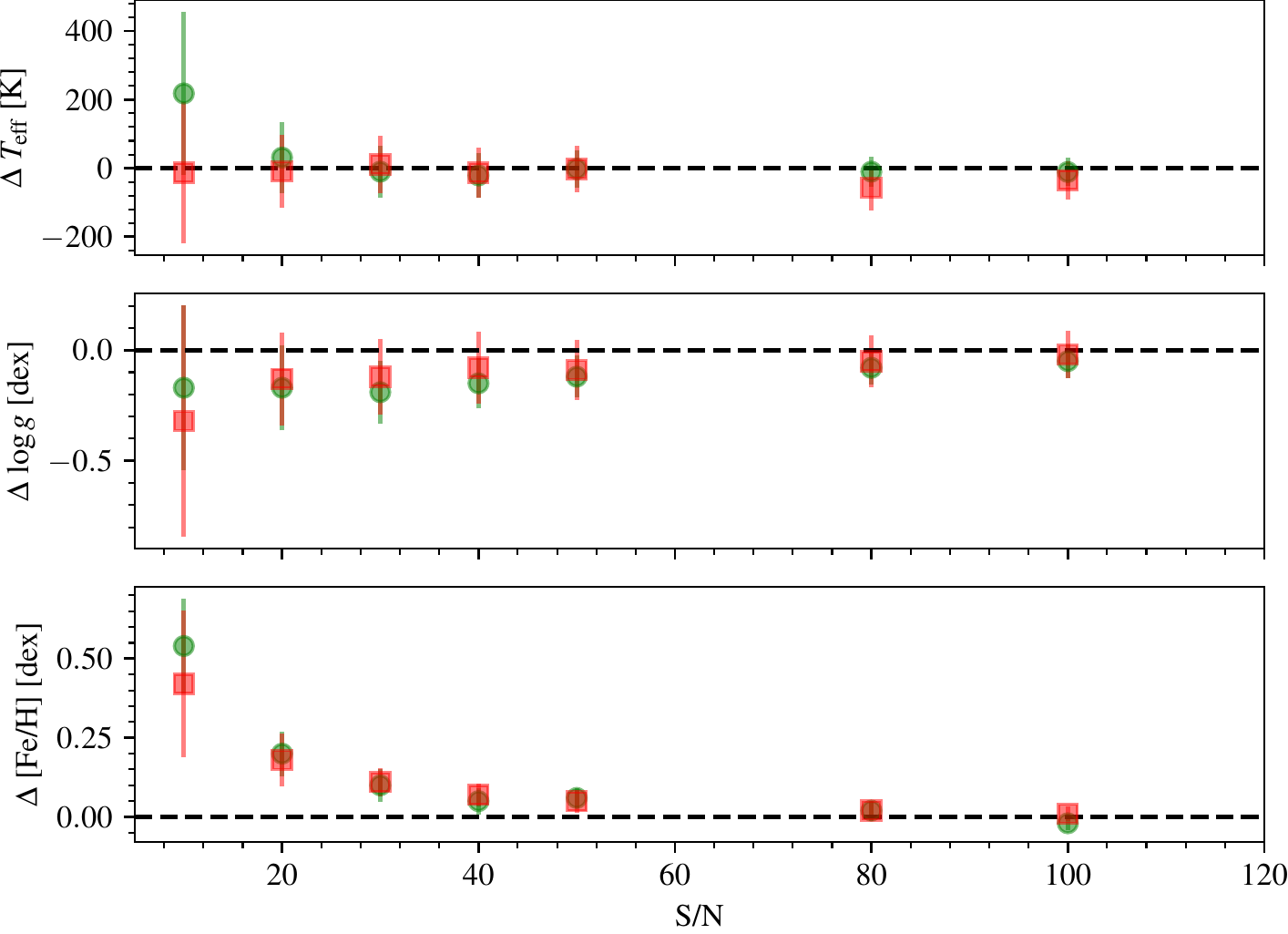}
    \includegraphics[width=0.45\textwidth]{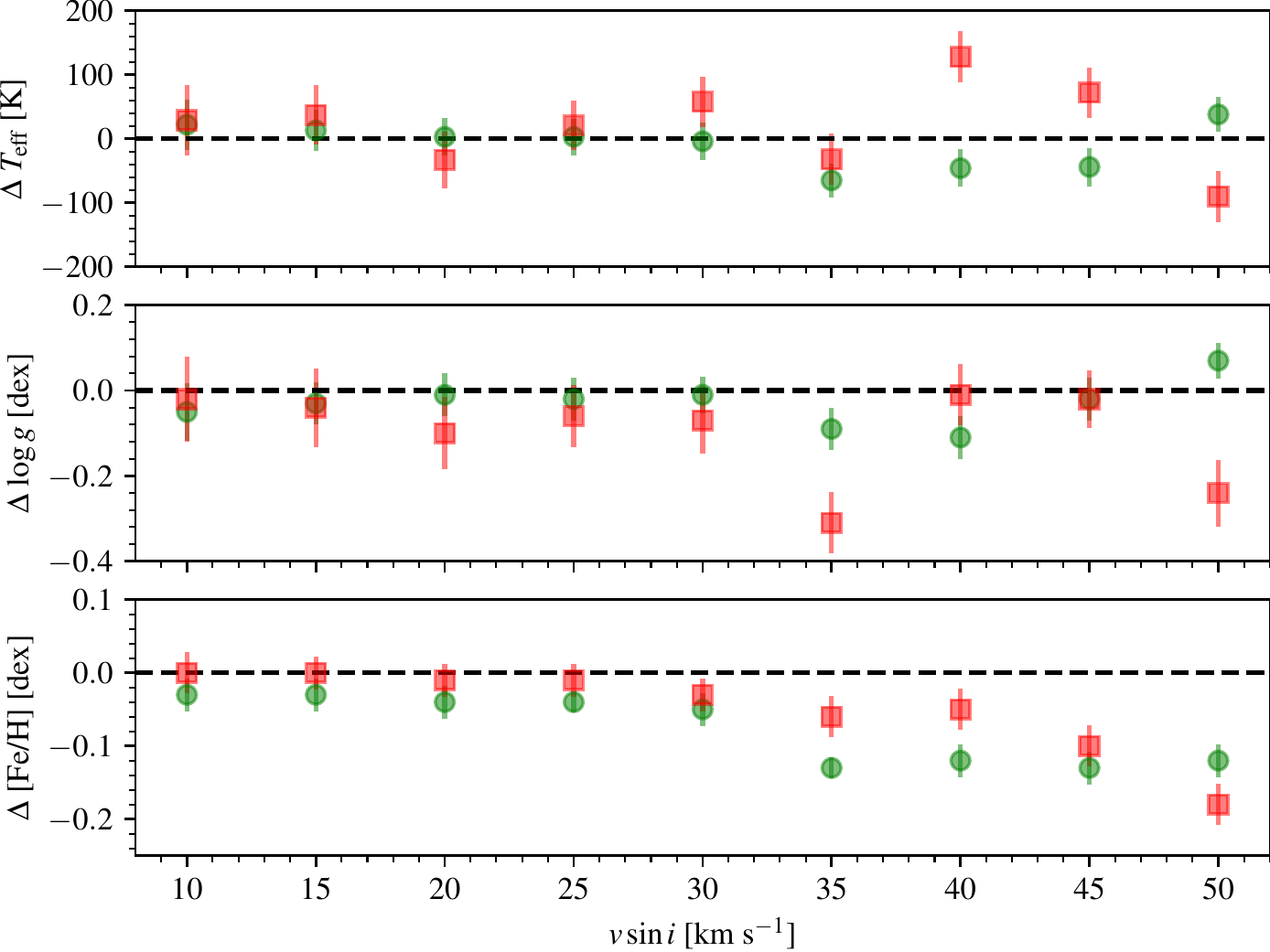}
    \caption{Variation of the stellar atmospheric parameters as a function of S/N (top panel) and   $\varv \sin{i}$ (bottom panel). The values corresponding to 18~Sco (G2~V) are represented by green circles, whereas those corresponding to $\epsilon$~Eri (K2~V) are displayed as red squares (see text for details).}
    \label{test_vsini_sn}
\end{figure}

Finally, we want to address the limitations of the {\sc SterParSyn} code regarding the analysis of late-type spectra under the spectral synthesis method. First, the observational data must be of high enough quality to be analysed. This means that, since low signal-to-noise ratios (S/N) translate into higher uncertainties on the stellar atmospheric parameters \citep{smi14}.  To compute the performance of {\sc SteParSyn} at different S/N, we took the spectra of 18~Sco and $\epsilon$~Eri and we degraded them to the following S/N values:  10, 20, 30, 40, 50, 80, and 100. Then, we ran {\sc SteParSyn} in order to calculate the stellar parameters of the degraded spectra. According to these calculations, the parameters have uncertainties at the level of 100~K for $T_{\rm eff}$, 0.2~dex for $\log{g}$, and 0.1~dex for [Fe/H] at S/N~$=$~20. (see Fig.~\ref{test_vsini_sn}). In addition, at S/N~$=$~10 the errors on the stellar parameters reach the level of 200~K, 0.5~dex, and 0.2~dex. In light of these numbers, the stellar parameters become unreliable for those spectra with S/N~$<$~20. Second, {\sc SteParSyn} cannot analyse double-lined spectra because it is not capable of disentangling two or more stellar components. A third limitation is the $\varv \sin{i}$ of the star under analysis. The fastest rotator in our sample is $\theta$~Boo~A with a $\varv \sin{i}$ of $\approx$~30~km~s$^{-1}$ and we were able to calculate its stellar parameters with {\sc SteParSyn}. Therefore, the limit on $\varv \sin{i}$ for the code should be above 30~km~s$^{-1}$. In fact, the limit on $\varv \sin{i}$ is related to the size of the spectral regions under analysis \citep{tsa14}. In general terms, the stellar rotation dilutes the spectral lines producing both narrow, broad, and likely blended spectral features. The line dilution should translate into less reliable stellar parameters, which we should be able to quantify. To that aim, we broadened the spectra of the stars 18~Sco and $\epsilon$~Eri by means of the Gray rotation kernel \citep{gra08} to different $\varv \sin{i}$ values. We chose these $\varv \sin{i}$ values to be in the 10--50~km~s$^{-1}$ range, taking a step of 5~km~s$^{-1}$. Thus, we calculated the stellar parameters of these rotationally-broadened spectra with {\sc SteParSyn}. Then, we compared this value to the parameters obtained for the original unbroadened spectra of both 18~Sco and $\epsilon$~Eri. We represent these differences as functions of  $\varv \sin{i}$ in Fig.~\ref{test_vsini_sn}. In all, {\sc SteParSyn} can recover the stellar parameters up to $\varv \sin{i}$~$=$~30~km~s$^{-1}$ with differences of no more than 50~K, 0.1~dex, and 0.05~dex for $T_{\rm eff}$, $\log{g}$, and [Fe/H], respectively.  However, {\sc SteParSyn} does not recover the expected stellar parameters for $\varv \sin{i}$~$\geq$~35~km~s$^{-1}$.  

\begin{table*}[ht]

\caption{Summary of the MC simulations on $T_{\rm eff}$, $\log{g}$, and [Fe/H] calculated with {\sc SteParSyn} against three samples of stellar parameters: literature values, {\sc StePar}, and the trigonometric surface gravities derived using PARAM. We provide differences on each parameter alongside the Pearson ($r_p$) and the Spearman ($r_s$) correlation coefficients (see text for details).}
\label{mc_table}
\centering
\begin{tabular}{lcccc}
\hline\hline\noalign{\smallskip}
Sample & Parameter & Difference & $r_p$ & $r_s$ \\
\hline\noalign{\smallskip}
& $T_{\rm eff}$          & $-$12~$\pm$~117~K   &  0.19~$\pm$~0.10  &  0.21~$\pm$~0.10 \\
Literature & $\log{g}$              & 0.04~$\pm$~0.14~dex & $-$0.34~$\pm$~0.11  &  0.04~$\pm$~0.14 \\
& $\lbrack$Fe/H$\rbrack$ & 0.05~$\pm$~0.09~dex &  $-$0.04~$\pm$~0.16  &  $-$0.05~$\pm$~0.16 \\ 
\hline
\noalign{\smallskip}
& $T_{\rm eff}$          & 40~$\pm$~87~K   &  0.07~$\pm$~0.12  &  0.06~$\pm$~0.13 \\
{\sc StePar} & $\log{g}$              & $-$0.04~$\pm$~0.21~dex & $-$0.51~$\pm$~0.13 &  $-$~0.41~$\pm$~0.12 \\
& $\lbrack$Fe/H$\rbrack$ & 0.04~$\pm$~0.08~dex &  0.28~$\pm$~0.12  &  0.21~$\pm$~0.11\\
\hline
\noalign{\smallskip}
& $T_{\rm eff}$          & ---  &  ---  &  --- \\
PARAM$^a$ & $\log{g}$        & $-$0.03~$\pm$~0.11~dex & $-$0.12~$\pm$~0.12  &  0.24~$\pm$~0.12 \\
& $\lbrack$Fe/H$\rbrack$ & --- &  ---  & --- \\
\hline

\end{tabular}
\tablefoot{
\tablefoottext{a}{PARAM only provides $\log{g}$ values.}
}
\end{table*}
\section{Conclusions}
\label{sec:conclusions}

In this work, we provide a description of the {\sc SteParSyn} code. The code is designed to infer the atmospheric parameters ($T_{\rm eff}$, $\log{g}$, [Fe/H], and $V_{\rm broad}$) of FGKM-type stars under the spectral synthesis method. The {\sc SteParSyn} code is publicly available to the community in a {\tt GitHub} repository. In summary,  it relies on the  MCMC sampler {\tt emcee} in conjunction with an spectral emulator that can interpolate spectra down to a precision $<$~1\%.  We also provide a grid of synthetic spectra with the code that allow the user to characterise the spectra of FGKM-type stars with parameters in the range of 3\,500 to 7\,000~K in $T_{\rm eff}$, 0.0 to 5.5~dex in $\log{g}$, and $-$2.0 to 1.0~dex in [Fe/H].\\

We tested the performance of the code against a sample of well-known 35 FGKM-type stars observed with the HERMES spectrograph.  We found that {\sc SteParSyn} infers stellar parameters that are consistent with the reference values. In addition, the $\log{g}$ values inferred with {\sc SteParSyn} are compatible with the trigonometric surface gravities provided by PARAM down to the 0.1~dex mark. Furthermore, we found good agreement between the results provided by both {\sc SteParSyn} and {\sc StePar}.\\

Finally, we want to note the limitations of the synthetic method that should be taken into consideration when using {\sc SteParSyn}. First, the spectroscopic data must have an S/N~$\ge$~20. Second, the code is meant to analyse only single-lined spectra as it cannot disentangle the flux of two or more stellar components. Third, the implementation of the spectral synthesis method provided in this work can only give reliable parameters for stars with $\varv \sin{i}$ on or below the 30~km~s$^{-1}$ mark.

\begin{acknowledgements}
 We would like to thank the anonymous referee for his/her comments and suggestions that helped to improve the paper. We acknowledge financial support from the Agencia Estatal de Investigaci\'on of the Ministerio de Ciencia, Innovaci\'on y Universidades through projects PID2019-109522GB-C51,54/AEI/10.13039/501100011033. HMT and JIGH acknowledge financial support from the Centre of Excellence "Severo Ochoa" and "Mar\'{i}a de Maeztu" awards to the Instituto de Astrof\'{i}sica de Canarias (SEV-2015-0548) and Centro de Astrobiolog\'{i}a (MDM-2017-0737). JIGH also acknowledges financial support from the Spanish Ministry of Science and Innovation (MICINN) project AYA2017-86389-P, and also from the Spanish MICINN under 2013 Ram\'on y Cajal program RYC-2013-14875. E.M. acknowledges financial support from the Spanish Ministerio de Universidades through fellowship FPU15/01476. This work is based on observations obtained with the HERMES spectrograph, which is supported by the Research Foundation - Flanders (FWO), Belgium, the Research Council of KU Leuven, Belgium, the Fonds National de la Recherche Scientifique (F.R.S.-FNRS), Belgium, the Royal Observatory of Belgium, the Observatoire de Genève, Switzerland and the Thüringer Landessternwarte Tautenburg, Germany. This research has made use of the SIMBAD database, operated at CDS, Strasbourg, France. This work has made use of data from the European Space Agency (ESA) mission {\it Gaia} (\url{https://www.cosmos.esa.int/gaia}), processed by the {\it Gaia} Data Processing and Analysis Consortium (DPAC, \url{https://www.cosmos.esa.int/web/gaia/dpac/consortium}). Funding for the DPAC has been provided by national institutions, in particular the institution participating in the {\it Gaia} Multilateral Agreement. This research made use of Astropy,\footnote{\url{http://www.astropy.org}} a community-developed core Python package for Astronomy \citep{astropy:2013, astropy:2018}.  
 
\end{acknowledgements}
\bibliographystyle{aa}
\bibliography{SteParSyn}
\begin{appendix}
\section{Extra material}
 \label{appendix}

{The sample analysed in this work is given in Table~\ref{obslog}, whereas Table~\ref{reference_parameters} contains the reference stellar atmospheric parameters compiled from literature sources. The stellar parameters of those stars analysed with {\sc SteParSyn} are given in Table~\ref{steparsyn_results}, whereas Table~\ref{stepar_results} contains the {\sc StePar} parameters. Finally, the wavelength regions containing the \ion{Fe}{i,ii} lines studied in this work are given in Table~\ref{linelist}.}

\begin{table*}
\caption{Names, coordinates, visual magnitudes ($V$), spectral types (SpT), signal-to-noise ratios (S/N), radial velocities (RV), and parallaxes ($\pi$) for the stars analysed in this work. The parallaxes were collected from the {\it Gaia} EDR3 \citep{EDR3} unless stated otherwise.
 }\label{obslog}
\centering
\begin{tabular}{lrccccccc}

\hline\hline\noalign{\smallskip}
Name & HIP & $\alpha$ (J2000) & $\delta$ (J2000) & $V$ & SpT & S/N$^a$ & RV  & $\pi$  \\
     &     &                  &                  &  [mag]   &    &        &  [km s$^{-1}$]  & [mas]\\
\hline\noalign{\smallskip}
Sun (Vesta)        &    ... &      ...    &   ...        & ...  & G2~V        &  139 & ... &  ... \\
HD~4628            &   3765 & 00:48:22.98 & +05:16:50.21 & 5.74 & K2~V        &  122 &  $-$9.997~$\pm$~0.010 &  134.50~$\pm$~0.058\\ 
$\eta$~Cas~A       &   3821 & 00:49:06.29 & +57:48:54.67 & 3.44 & G0~V        &  268 &     8.547~$\pm$~0.015 &  168.83~$\pm$~0.17\\ 
$\mu$~Cas          &   5336 & 01:08:16.40 & +54:55:13.23 & 5.17 & G5~Vb       &  194 & $-$96.540~$\pm$~0.015 &  130.29~$\pm$~0.44\\
$\tau$~Cet         &   8102 & 01:44:04.08 & $-$15:56:14.93 & 3.50 & G8.5~V      &  194 & $-$16.530~$\pm$~0.011 &  273.81~$\pm$~0.17\\
HD~16160           &  12114 & 02:36:04.90 & +06:53:12.43 & 5.83 & K3~V        &  187 &    25.763~$\pm$~0.009 &  138.34~$\pm$~0.32\\ 
$\theta$~Per~A     &  12777 & 02:44:11.99 & +49:13:42.41 & 4.11 & F7~V        &  250 &    24.620~$\pm$~0.028 &   89.69~$\pm$~0.16\\ 
$\alpha$~Cet       &  14135 & 03:02:16.77 & +04:05:23.06 & 2.53 & M1.5~IIIa   &  117 & $-$25.784~$\pm$~0.011 &     13.09~$\pm$~0.44\\
$\epsilon$~Eri     &  16537 & 03:32:55.85 & $-$09:27:29.73 & 3.73 & K2~V        &  208 &    16.449~$\pm$~0.010 &  310.58~$\pm$~0.14\\
HD~22879           &  17147 & 03:40:22.07 & $-$03:13:01.13 & 6.67 & F9~V        &  112 &   120.475~$\pm$~0.023 &   38.325~$\pm$~0.031\\
$\delta$~Eri       &  17378 & 03:43:14.90 & $-$09:45:48.21 & 3.54 & K1~III-IV   &  190 &  $-$6.155~$\pm$~0.008 &  110.03~$\pm$~0.19\\
Aldebaran          &  21421 & 04:35:55.24 & +16:30:33.49 & 0.86 & K5~III      &  151 &    54.120~$\pm$~0.009 &  47.253~$\pm$~0.096\\
$\pi^3$~Ori        &  22449 & 04:49:50.41 & +06:57:40.59 & 3.19 & F6~V        &   93 &    25.074~$\pm$~0.070 & 124.62~$\pm$~0.23\\ 
HD~49933           &  32851 & 06:50:49.83 & $-$00:32:27.18 & 5.77 & F2~V        &  130 & $-$12.275~$\pm$~0.054 &  33.534~$\pm$~0.043\\
Procyon            &  37279 & 07:39:18.12 & +05:13:29.96 & 0.37 & F5~IV-V     &  432 &  $-$5.275~$\pm$~0.023 &   284.56~$\pm$~1.26$^b$\\
$\beta$~Gem        &  37826 & 07:45:18.95 & +28:01:34.32 & 1.14 & K0~IIIb     &  183 &     3.536~$\pm$~0.007 &   96.54~$\pm$~0.27$^b$\\
$\mu$~Leo          &  48455 & 09:52:45.82 & +26:00:25.03 & 3.88 & K2~III      &  112 &    13.764~$\pm$~0.007 &  26.10~$\pm$~0.20 \\
20~LMi             &  49081 & 10:01:00.66 & +31:55:25.22 & 5.40 & G3~V        &  224 &    56.118~$\pm$~0.010 &   66.996~$\pm$~0.092\\
36~UMa             &  51459 & 10:30:37.58 & +55:58:49.94 & 4.72 & F8~V        &  193 &     8.880~$\pm$~0.016 &   77.249~$\pm$~0.081\\
$\beta$~Vir        &  57757 & 11:50:41.72 & +01:45:52.99 & 3.60 & F9~V        &  152 &     4.757~$\pm$~0.014 &   90.90~$\pm$~0.19 \\
Gmb~1830           &  57939 & 11:52:58.77 & +37:43:07.24 & 6.45 & G8~Vp       &  205 & $-$98.001~$\pm$~0.022 &  109.030~$\pm$~0.020 \\
HD~107328          &  60172 & 12:20:20.98 & +03:18:45.26 & 4.96 & K0~IIIb     &  145 &    36.656~$\pm$~0.009 &    9.67~$\pm$~0.15\\
$\epsilon$~Vir     &  63608 & 13:02:10.60 & +10:57:32.94 & 2.79 & G8~III      &  204 & $-$14.169~$\pm$~0.009 &   30.21~$\pm$~0.19         \\
$\beta$~Com        &  64394 & 13:11:52.39 & +27:52:41.45 & 4.25 & G0~V        &  231 &     5.438~$\pm$~0.016 &  108.73~$\pm$~0.17\\ 
$\eta$~Boo         &  67927 & 13:54:41.08 & +18:23:51.80 & 2.68 & G0~IV       &  258 &     7.109~$\pm$~0.028 &     87.75~$\pm$~1.24$^b$\\
Arcturus           &  69673 & 14:15:39.67 & +19:10:56.67 &$-$0.05 & K1.5~III  &  207 &    $-$4.880~$\pm$~0.009 &   88.83~$\pm$~0.54$^b$\\
$\theta$~Boo~A &  70497 & 14:25:11.80 & +51:51:02.68 & 4.05 & F7~V      & 223 & $-$10.734~$\pm$~0.117 &   69.07~$\pm$~0.16\\
18~Sco             &  79672 & 16:15:37.27 & $-$08:22:09.98 & 5.50 & G2~Va       &  202 &    12.000~$\pm$~0.011 & 70.737~$\pm$~0.063\\
$\gamma$~Sge       &  98337 & 19:58:45.43 & +19:29:31.73 & 3.47 & M0~III      &  127 &   $-$33.712~$\pm$~0.008 & 13.09~$\pm$~0.44\\
HD~189733          &  98505 & 20:00:43.71 & +22:42:39.07 & 7.65 & K2~V        &  153 & $-$1.925~$\pm$~0.011 &   50.567~$\pm$~0.016\\
61~Cyg~A           & 104214 & 21:06:53.94 & +38:44:57.90 & 5.21 & K5~V        &  138 &   $-$65.562~$\pm$~0.013 & 285.995~$\pm$~0.060\\
61~Cyg~B           & 104217 & 21:06:55.26 & +38:44:31.36 & 6.03 & K7~V        &  115 &   $-$64.232~$\pm$~0.019 & 286.005~$\pm$~0.029\\
HD~209458          & 108859 & 22:03:10.77 & +18:53:03.55 & 7.63 & F9~V        &  167 & $-$14.678~$\pm$~0.017 &   20.769~$\pm$~0.027\\
$\xi$~Peg          & 112447 & 22:46:41.58 & +12:10:22.39 & 4.20 & F7~V        &  296 & $-$5.841~$\pm$~0.030  &   60.92~$\pm$~0.17\\
HD~220009          & 115227 & 23:20:20.58 & +05:22:52.70 & 5.07 & K2~III      &  213 &   40.793~$\pm$~0.009 & 9.09~$\pm$~0.12\\
\hline
\end{tabular}
\newline
\tablefoot{
\tablefoottext{a}{Computed as in \citet{sou15}.}
\tablefoottext{b}{Parallax from {\it Hipparcos} \citep{newhip}.}
}
\end{table*}

\begin{table*}[ht]
\caption{Reference stellar atmospheric parameters for the stars analysed in this work.}
\label{reference_parameters}
\centering
\begin{tabular}{lcccccccc}
\hline\hline\noalign{\smallskip}
Name           & SpT        &  $T_{\rm eff}$ & $\log{g}$       & Ref.$^a$   & [Fe/H]             & Ref.$^b$ & $\varv \sin{i}$  & Ref.$^c$\\
               &            &       [K]      &  [dex]          &            & [dex]              &          &  [km~s$^{-1}$]   &         \\
\hline\noalign{\smallskip}
Sun            &  G2~V      & 5771~$\pm$~1   & 4.44~$\pm$~0.01 & Hei15      &    0.03~$\pm$~0.05 & Jof14    &            1.6      & Vf05         \\
HD~4628        &  K2~V      & 4950~$\pm$~14  & 4.64~$\pm$~0.01 & Boy12b$^d$ & $-$0.27~$\pm$~0.04 & Pastel   &              2.0    &  Mar14 \\ 
$\eta$~Cas~A   &  G0~V      & 6003~$\pm$~24  & 4.39~$\pm$~0.01 & Boy12a$^d$ & $-$0.27~$\pm$~0.07 & Pastel   &       5.4           & Luc17          \\ 
$\mu$~Cas      &  G5~Vb     & 5308~$\pm$~29  & 4.41~$\pm$~0.06 & Hei15      & $-$0.81~$\pm$~0.03 & Jof14    &       1.1           &  Hal94       \\
$\tau$~Cet     &  G8.5~V    & 5414~$\pm$~21  & 4.49~$\pm$~0.03 & Hei15      & $-$0.49~$\pm$~0.03 & Jof14    &        1.6          & Jen11           \\
HD~16160       &  K3~V      & 4662~$\pm$~17  & 4.52~$\pm$~0.01 & Boy12b$^d$ & $-$0.13~$\pm$~0.06 & Pastel   &   1.3               & Que98          \\
$\theta$~Per~A &  F7~V      & 6157~$\pm$~37  & 4.26~$\pm$~0.01 & Boy12a$^d$ &    0.00~$\pm$~0.08 & Pastel   &          10.2        &   Luc17        \\
$\alpha$~Cet   &  M1.5~IIIa & 3796~$\pm$~65  & 0.68~$\pm$~0.23 & Hei15      & $-$0.45~$\pm$~0.47 & Jof14    &            6.9     &  Mas08         \\
$\epsilon$~Eri &  K2~V      & 5076~$\pm$~30  & 4.61~$\pm$~0.03 & Hei15      & $-$0.09~$\pm$~0.06 & Jof14    &      6.2            & Bud19     \\ 
HD~22879       &  F9~V      & 5868~$\pm$~89  & 4.27~$\pm$~0.04 & Hei15      & $-$0.86~$\pm$~0.05 & Jof14    &             4.2     &  Bud19         \\
$\delta$~Eri   &  K1~III-IV & 4954~$\pm$~30  & 3.76~$\pm$~0.02 & Hei15      &    0.06~$\pm$~0.05 & Jof14    &           3.2       &     Sot18     \\
Aldebaran      &  K5~III    & 3927~$\pm$~40  & 1.11~$\pm$~0.19 & Hei15      & $-$0.37~$\pm$~0.17 & Jof14    &        4.3          &  Mas08        \\
$\pi^3$~Ori    &  F6~V      & 6516~$\pm$~19  & 4.31~$\pm$~0.01 & Boy12a$^d$ &    0.05~$\pm$~0.07 & Pastel   &               18.5   & Luc17          \\ 
HD~49933       &  F2~V      & 6635~$\pm$~91  & 4.20~$\pm$~0.03 & Hei15      & $-$0.41~$\pm$~0.08 & Jof14    &           5.0       &  Tak05         \\
Procyon        &  F5~IV-V   & 6554~$\pm$~84  & 4.00~$\pm$~0.02 & Hei15      &    0.01~$\pm$~0.08 & Jof14    &             7.4     & Luc17          \\
$\beta$~Gem    &  K0~IIIb   & 4858~$\pm$~60  & 2.90~$\pm$~0.08 & Hei15      &    0.13~$\pm$~0.16 & Jof14    &               2.8   & Mas08          \\
$\mu$~Leo      &  K2~III    & 4474~$\pm$~60  & 2.51~$\pm$~0.11 & Hei15      &    0.25~$\pm$~0.15 & Jof14    &           4.5       &  Mas08     \\
20~LMi         &  G3~V      & 5612~$\pm$~52  & 4.26~$\pm$~0.03 & Boy12a$^d$ &    0.18~$\pm$~0.12 & Pastel   &             4.7     &  Luc17          \\
36~UMa         &  F8~V      & 6233~$\pm$~68  & 4.41~$\pm$~0.03 & Boy12a$^d$ & $-$0.13~$\pm$~0.05 & Pastel   &           5.3       & Luc17          \\
$\beta$~Vir    &  F9~V      & 6083~$\pm$~41  & 4.10~$\pm$~0.02 & Hei15      &    0.24~$\pm$~0.07 & Jof14    &          6.4        &    Luc17      \\
Gmb~1830       &  G8~Vp     & 4827~$\pm$~55  & 4.60~$\pm$~0.03 & Hei15      & $-$1.46~$\pm$~0.39 & Jof14    &   0.0               &     Lat02     \\
HD107328       &  K0~IIIb   & 4496~$\pm$~59  & 2.09~$\pm$~0.13 & Hei15      & $-$0.33~$\pm$~0.16 & Jof14    &           1.9       &  Mas08       \\
$\epsilon$~Vir &  G8~III    & 4983~$\pm$~61  & 2.77~$\pm$~0.02 & Hei15      &    0.15~$\pm$~0.16 & Jof14    &             1.4     & Jof15          \\
$\beta$~Com    &  G0~V      & 5936~$\pm$~33  & 4.37~$\pm$~0.01 & Boy12a$^d$ &    0.07~$\pm$~0.09 & Pastel   &                 6.6 & Luc17          \\ 
$\eta$~Boo     &  G0~IV     & 6099~$\pm$~28  & 3.79~$\pm$~0.02 & Hei15      &    0.32~$\pm$~0.08 & Jof14    &               14.4   & Luc17          \\
Arcturus       &  K1.5~III  & 4286~$\pm$~35  & 1.60~$\pm$~0.20 & Hei15      & $-$0.52~$\pm$~0.08 & Jof14    &               4.2   & Mas08          \\
$\theta$~Boo~A &  F7~V      & 6265~$\pm$~41  & 4.05~$\pm$~0.04 & Boy12a$^d$ & $-$0.03~$\pm$~0.03 & Pastel   &         30.4         & Luc17          \\
18~Sco         &  G2~Va     & 5810~$\pm$~80  & 4.44~$\pm$~0.03 & Hei15      &    0.03~$\pm$~0.03 & Jof14    &                4.4  &  Luc17         \\
$\gamma$~Sge   &  M0~III    & 3807~$\pm$~49  & 1.05~$\pm$~0.32 & Hei15      & $-$0.17~$\pm$~0.39 & Jof14    &                 5.8 & Mas08          \\
HD~189733      &  K2~V      & 4875~$\pm$~43  & 4.56~$\pm$~0.03 & Boy15$^d$  & $-$0.02~$\pm$~0.04 & Pastel   &               4.5   & Luc17          \\
61~Cyg~A       &  K5~V      & 4374~$\pm$~22  & 4.63~$\pm$~0.04 & Hei15      & $-$0.33~$\pm$~0.38 & Jof14    &           1.9       &    Que98 \\
61~Cyg~B       &  K7~V      & 4044~$\pm$~32  & 4.67~$\pm$~0.04 & Hei15      & $-$0.38~$\pm$~0.03 & Jof14    &           0.7       &    Que98      \\
HD~209458      &  F9~V      & 6092~$\pm$~103 & 4.28~$\pm$~0.10 & Boy15$^d$  &    0.03~$\pm$~0.05 & Pastel   &               3.9   & Sot18          \\
$\xi$~Peg      &  F7~V      & 6168~$\pm$~36  & 3.95~$\pm$~0.01 & Boy12a$^d$ & $-$0.27~$\pm$~0.08 & Pastel   &          9.7        & Luc17         \\
HD~220009      &  K2~III    & 4217~$\pm$~60  & 1.43~$\pm$~0.12 & Hei15      & $-$0.74~$\pm$~0.13 & Jof14    &                 1.2  &    Med99       \\
\hline
\end{tabular}
\newline
\tablefoot{
\tablefoottext{a}{$T_{\rm eff}$ and $\log{g}$ reference. Hei15: \citet{hei15a}; Boy12a: \citet{boyafg}; Boy12b: \citet{boykm};  Boy15: \citet{boy2pla}.}
\tablefoottext{b}{[Fe/H] reference. Jof14: \citet{jof14}; Pastel: Literature average computed from the PASTEL catalogue \citep{soub16}.}
\tablefoottext{c}{$\varv \sin{i}$ reference. Hal94: \citet{mucasvsini}; Que98: \citet{61cygvsini}; Med99: \citet{Me99}; Lat02: \citet{gmbvsini}; Tak05: \citet{tak05}; Vf05: \citet{valfis05}; Mas08: \citet{mas08}; Jen11: \citet{taucetvsini}; Mar14: \citet{hd4628vsini}; Jof15: \citet{ejof15}; Luc17: \citet{luc17}; Sot18: \citet{sot18}; Bud19: \citet{bud19}.}
\tablefoottext{d}{$\log{g}$ values were computed from the corresponding literature values of mass and radius.}
}
\end{table*}

\begin{table*}[ht]
\caption{Stellar atmospheric parameters derived with {\sc SteParSyn} ($T_{\rm eff}$, $\log{g}$, [Fe/H], and $V_{\rm broad}$), trigonometric gravities ($\log{g_{\rm trig}}$), mass ($M_{*}$), and radius ($R_{*}$) for  the stars analysed in this work.}
\label{steparsyn_results}
\centering
\begin{tabular}{lccccccc}
\hline\hline\noalign{\smallskip}
Name & $T_{\rm eff}$ & $\log{g}$ & [Fe/H]  & $V_{\rm broad}$ & $\log{g_{\rm trig}}$ &  $M_{*}$ & $R_{*}$ \\
     & [K]           & [dex]     & [dex]   & [km\,s$^{-1}$]    &  [dex] & [$M_\odot$]  & [$R_\odot$]\\
\hline\noalign{\smallskip}
Sun            & 5775~$\pm$~13 & 4.41~$\pm$~0.02 & $-$0.04~$\pm$~0.01 &  3.71~$\pm$~0.03 &  ...          &        1        &  1 \\
HD~4628        & 4969~$\pm$~37 & 4.57~$\pm$~0.05 & $-$0.33~$\pm$~0.02 & 2$^a$            & 4.59~$\pm$~0.02 & 0.720~$\pm$~0.014 &  0.691~$\pm$~0.007\\
$\eta$~Cas~A   & 5874~$\pm$~34 & 4.30~$\pm$~0.06 & $-$0.33~$\pm$~0.02 &  4.00~$\pm$~0.07 & 4.30~$\pm$~0.02 & 0.881~$\pm$~0.014 &  1.062~$\pm$~0.022\\
$\mu$~Cas      & 5370~$\pm$~35 & 4.58~$\pm$~0.05 & $-$0.87~$\pm$~0.02 &  1.1$^a$         & 4.50~$\pm$~0.01 & 0.747~$\pm$0.008  &  0.781~$\pm$~0.018\\
$\tau$~Cet     & 5400~$\pm$~60 & 4.58~$\pm$~0.05 & $-$0.53~$\pm$~0.04 &  1.6$^a$         & 4.54~$\pm$~0.02 & 0.760~$\pm$~0.017 &  0.750~$\pm$~0.015\\
HD~16160       & 4900~$\pm$~45 & 4.65~$\pm$~0.07 & $-$0.19~$\pm$~0.02 & 1.3$^a$          & 4.60~$\pm$~0.02 & 0.741~$\pm$~0.015 &  0.691~$\pm$~0.006\\
$\theta$~Per~A & 6302~$\pm$~20 & 4.25~$\pm$~0.03 & $-$0.02~$\pm$~0.01 &  9.39~$\pm$~0.04 & 4.30~$\pm$~0.01 & 1.201~$\pm$~0.010 &  1.250~$\pm$~0.020\\
$\alpha$~Cet   & 3883~$\pm$~41 & 1.56~$\pm$~0.17 & $-$0.01~$\pm$~0.09 &  6.30~$\pm$~0.23 & 1.08~$\pm$~0.06 & 2.803~$\pm$~0.219 & 77.347~$\pm$~5.120\\
$\epsilon$~Eri & 5086~$\pm$~33 & 4.62~$\pm$~0.06 & $-$0.08~$\pm$~0.02 &  3.11~$\pm$~0.14 & 4.58~$\pm$~0.02 & 0.806~$\pm$~0.019 &  0.734~$\pm$~0.015\\
HD~22879       & 5763~$\pm$~30 & 4.13~$\pm$~0.05 & $-$0.91~$\pm$~0.02 &  3.97~$\pm$~0.08 & 4.29~$\pm$~0.01 & 0.977~$\pm$~0.015 &  1.135~$\pm$~0.023\\
$\delta$~Eri   & 4934~$\pm$~41 & 3.62~$\pm$~0.09 &    0.04~$\pm$~0.02 &  3.27~$\pm$~0.11 & 3.67~$\pm$~0.04 & 1.094~$\pm$~0.047 &  2.450~$\pm$~0.070\\
Aldebaran      & 3858~$\pm$~21 & 0.99~$\pm$~0.10 & $-$0.36~$\pm$~0.07 &  4.93~$\pm$~0.13 & 1.02~$\pm$~0.04 & 0.912~$\pm$~0.063 & 47.535~$\pm$~1.427\\
$\pi^3$~Ori    & 6358~$\pm$~34 & 4.04~$\pm$~0.08 & $-$0.02~$\pm$~0.02 & 18.20~$\pm$~0.14 & 4.25~$\pm$~0.01 & 1.230~$\pm$~0.012 &  1.339~$\pm$~0.029\\
HD~49933       & 6725~$\pm$~26 & 4.20~$\pm$~0.05 & $-$0.39~$\pm$~0.02 & 11.10~$\pm$~0.08 & 4.21~$\pm$~0.01 & 1.188~$\pm$~0.014 &  1.379~$\pm$~0.009\\
Procyon        & 6631~$\pm$~34 & 3.97~$\pm$~0.05 & $-$0.02~$\pm$~0.02 &  6.18~$\pm$~0.03 & 4.00~$\pm$~0.01 & 1.480~$\pm$~0.010 &  1.965~$\pm$~0.043\\
$\beta$~Gem    & 4756~$\pm$~23 & 2.76~$\pm$~0.07 &    0.02~$\pm$~0.03 &  4.08~$\pm$~0.07 & 2.73~$\pm$~0.03 & 1.819~$\pm$~0.075 &  9.324~$\pm$~0.215\\
$\mu$~Leo      & 4480~$\pm$~43 & 2.07~$\pm$~0.13 &    0.21~$\pm$~0.07 &  4.32~$\pm$~0.12 & 2.41~$\pm$~0.07 & 1.311~$\pm$~0.161 & 11.503~$\pm$~0.360\\
20~LMi         & 5691~$\pm$~32 & 4.27~$\pm$~0.06 &    0.13~$\pm$~0.02 &  4.04~$\pm$~0.07 & 4.29~$\pm$~0.02 & 1.010~$\pm$~0.016 &  1.158~$\pm$~0.021\\
36~UMa         & 6102~$\pm$~43 & 4.29~$\pm$~0.08 & $-$0.15~$\pm$~0.03 &  4.19~$\pm$~0.09 & 4.29~$\pm$~0.02 & 1.049~$\pm$~0.021 &  1.180~$\pm$~0.027\\
$\beta$~Vir    & 6024~$\pm$~28 & 3.88~$\pm$~0.04 &    0.04~$\pm$~0.02 &  5.24~$\pm$~0.05 & 4.03~$\pm$~0.02 & 1.226~$\pm$~0.041 &  1.722~$\pm$~0.032\\
Gmb~1830       & 4997~$\pm$~18 & 4.60~$\pm$~0.03 & $-$1.41~$\pm$~0.02 &  0$^a$           & 4.60~$\pm$~0.01 & 0.616~$\pm$~0.001 &  0.631~$\pm$~0.003\\
HD~107328      & 4427~$\pm$~20 & 1.62~$\pm$~0.06 & $-$0.51~$\pm$~0.16 &  5.00~$\pm$~0.07 & 1.81~$\pm$~0.08 & 1.049~$\pm$~0.180 & 20.428~$\pm$~0.540\\
$\epsilon$~Vir & 5019~$\pm$~23 & 2.73~$\pm$~0.05 &    0.08~$\pm$~0.03 &  5.16~$\pm$~0.06 & 2.72~$\pm$~0.01 & 2.878~$\pm$~0.020 & 11.939~$\pm$~0.206\\
$\beta$~Com    & 6000~$\pm$~22 & 4.40~$\pm$~0.04 &    0.10~$\pm$~0.04 &  5.56~$\pm$~0.05 & 4.40~$\pm$~0.02 & 1.130~$\pm$~0.024 &  1.075~$\pm$~0.025\\
$\eta$~Boo     & 5956~$\pm$~18 & 3.64~$\pm$~0.04 &    0.15~$\pm$~0.01 & 13.55~$\pm$~0.05 & 3.73~$\pm$~0.01 & 1.587~$\pm$~0.022 &  2.762~$\pm$~0.042\\
Arcturus       & 4294~$\pm$~24 & 1.49~$\pm$~0.04 & $-$0.59~$\pm$~0.02 &  4.98~$\pm$~0.07 & 1.58~$\pm$~0.04 & 0.887~$\pm$~0.043 & 24.457~$\pm$~0.606\\
$\theta$~Boo~A & 6300~$\pm$~23 & 3.93~$\pm$~0.04 & $-$0.09~$\pm$~0.01 & 29.50~$\pm$~0.12 & 4.06~$\pm$~0.01 & 1.255~$\pm$~0.039 &  1.661~$\pm$~0.010\\
18~Sco         & 5763~$\pm$~18 & 4.42~$\pm$~0.03 &   -0.02~$\pm$~0.01 &  3.80~$\pm$~0.03 & 4.39~$\pm$~0.02 & 0.977~$\pm$~0.012 &  1.017~$\pm$~0.023\\
$\gamma$~Sge   & 3943~$\pm$~25 & 1.27~$\pm$~0.10 &    0.00~$\pm$~0.15 &  5.35~$\pm$~0.14 & 1.23~$\pm$~0.08 & 1.446~$\pm$~0.263 & 46.623~$\pm$~2.241\\
HD~189733      & 4995~$\pm$~32 & 4.51~$\pm$~0.06 & $-$0.01~$\pm$~0.01 &  4.08~$\pm$~0.09 & 4.55~$\pm$~0.02 & 0.800~$\pm$~0.018 &  0.758~$\pm$~0.003\\
61~Cyg~A       & 4576~$\pm$~36 & 4.69~$\pm$~0.06 & $-$0.37~$\pm$~0.02 &  1.9$^a$         & 4.69~$\pm$~0.02 & 0.624~$\pm$~0.010 &  0.574~$\pm$~0.005\\
61~Cyg~B       & 4088~$\pm$~26 & 4.61~$\pm$~0.07 & $-$0.44~$\pm$~0.04 &  0.7$^a$         & 4.57~$\pm$~0.04 & 0.451~$\pm$~0.024 &  0.562~$\pm$~0.015\\
HD~209458      & 6028~$\pm$~24 & 4.21~$\pm$~0.04 & $-$0.07~$\pm$~0.01 &  5.55~$\pm$~0.05 & 4.29~$\pm$~0.01 & 1.054~$\pm$~0.012 &  1.178~$\pm$~0.027\\
$\xi$~Peg      & 6091~$\pm$~18 & 3.73~$\pm$~0.03 & $-$0.40~$\pm$~0.01 &  9.02~$\pm$~0.05 & 3.89~$\pm$~0.01 & 1.096~$\pm$~0.001 &  1.909~$\pm$~0.016\\
HD~220009      & 4260~$\pm$~26 & 1.57~$\pm$~0.05 & $-$0.79~$\pm$~0.03 & 4.42~$\pm$~0.07  & 1.58~$\pm$~0.01 & 0.777~$\pm$~0.003 & 22.920~$\pm$~0.111\\
\hline
\end{tabular}
\newline
\tablefoot{
\tablefoottext{a}{We fixed $V_{\rm broad}$ to  the values of  $\varv\sin{i}$ reported in the literature. These were taken from \citet{mucasvsini} for $\mu$~Cas, \citet{taucetvsini} for $\tau$~Cet, \citet{gmbvsini} for Gmb~1830, \citet{hd4628vsini} for HD~4628, and \citet{61cygvsini} for both the 61~Cyg system and HD~16160.}
}
\end{table*}

\begin{table*}[ht]
\caption{Stellar atmospheric parameters ($T_{\rm eff}$, $\log{g}$, [Fe/H], and $\xi$) for the stars analysed work using the {\sc StePar} code. Missing values correspond to those stars that cannot be analysed under the EW method.}
\label{stepar_results}
\centering
\begin{tabular}{lcccc}
\hline\hline\noalign{\smallskip}
Name & $T_{\rm eff}$ & $\log{g}$ & [Fe/H]  &  $\xi$ \\
     & [K]           & [dex]     & [dex]   & [km\,s$^{-1}$]\\
\hline\noalign{\smallskip}
Sun            & 5792~$\pm$~43 & 4.36~$\pm$~0.09  & 0.00~$\pm$~0.04 & 0.74~$\pm$~0.09\\
HD~4628        & 4999~$\pm$~71 & 4.40~$\pm$~0.17 &$-$0.33~$\pm$~0.05 & 0.55~$\pm$~0.28\\
$\eta$~Cas~A   & 5952~$\pm$~30 & 4.36~$\pm$~0.07 &$-$0.26~$\pm$~0.03 & 0.94~$\pm$~0.06\\
$\mu$~Cas      & 5297~$\pm$~40 & 4.22~$\pm$~0.11 & $-$0.89~$\pm$~0.03 & 0.57~$\pm$~0.11\\
$\tau$~Cet     & 5334~$\pm$~49 & 4.30~$\pm$~0.12 & $-$0.54~$\pm$~0.04 &0.48~$\pm$~0.15\\
HD~16160       & 4935~$\pm$~90 & 4.36~$\pm$~0.26 & $-$0.23~$\pm$~0.06 & 0.82~$\pm$~0.33 \\
$\theta$~Per~A & 6289~$\pm$~37 & 4.20~$\pm$~0.07 & 0.01~$\pm$~0.03  & 1.29~$\pm$~0.04 \\
$\alpha$~Cet   & -- & -- & -- & --  \\
$\epsilon$~Eri & 5096~$\pm$~68 & 4.45~$\pm$~0.16 & $-$0.11~$\pm$~0.05 & 0.73~$\pm$~0.24 \\
HD~22879       & 5780~$\pm$~32 & 4.08~$\pm$~0.08 & $-$0.90~$\pm$~0.03 & 0.85~$\pm$~0.06 \\
$\delta$~Eri   & 4984~$\pm$~79 & 3.60~$\pm$~0.17 &    0.08~$\pm$~0.07 &  0.72~$\pm$~0.21 \\
Aldebaran      & -- & -- & --  & -- \\
$\pi^3$~Ori    & -- & -- & -- & -- \\
HD~49933       & 6689~$\pm$~62 & 4.05~$\pm$~0.12 & $-$0.31~$\pm$~0.04 & 1.28~$\pm$~0.06 \\
Procyon        & 6604~$\pm$~33 & 3.75~$\pm$~0.06 & $-$0.04~$\pm$~0.02 & 1.57~$\pm$~0.04 \\
$\beta$~Gem    & 4883~$\pm$~65 & 2.86~$\pm$~0.22 &    0.17~$\pm$~0.09 & 0.99~$\pm$~0.18 \\
$\mu$~Leo      & 4529~$\pm$~94 & 2.73~$\pm$~0.30 &   0.39~$\pm$~0.06 & 1.26~$\pm$~0.10\\
20~LMi         & 5753~$\pm$~50 & 4.20~$\pm$~0.11 &   0.23~$\pm$~0.05  & 0.77~$\pm$~0.13 \\
36~UMa         & 6220~$\pm$~35 & 4.40~$\pm$~0.08 & $-$0.09~$\pm$~0.03 & 1.06~$\pm$~0.05 \\
$\beta$~Vir    & 6224~$\pm$~42 & 4.17~$\pm$~0.07 &   0.18~$\pm$~0.03 & 1.16~$\pm$~0.06 \\
Gmb~1830       & 5064~$\pm$~40 & 4.37~$\pm$~0.08 & $-$1.37~$\pm$~0.03 & 0.96~$\pm$~0.13 \\
HD~107328      & 4385~$\pm$~36 & 1.65~$\pm$~0.15 & $-$0.47~$\pm$~0.03 & 1.53~$\pm$~0.05 \\
$\epsilon$~Vir     & 5078~$\pm$~39 & 2.71~$\pm$~0.13 &    0.17~$\pm$~0.04 & 1.28~$\pm$~0.05 \\
$\beta$~Com    & 6041~$\pm$~39 & 4.32~$\pm$~0.09 &    0.07~$\pm$~0.03 & 0.90~$\pm$~0.07 \\
$\eta$~Boo     & 6196~$\pm$~55 & 3.72~$\pm$~0.13 &    0.36~$\pm$~0.04 & 1.61~$\pm$~0.06\\
Arcturus       & 4259~$\pm$~40 & 1.37~$\pm$~0.19 & $-$0.59~$\pm$~0.04 &  1.48~$\pm$~0.05 \\
$\theta$~Boo~A & -- & -- & -- & --\\
18~Sco         & 5799~$\pm$~39 & 4.35~$\pm$~0.08 &   0.02~$\pm$~0.03 & 0.81~$\pm$~0.07 \\
$\gamma$~Sge   & -- & -- & --  & --\\
HD~189733      & 5060~$\pm$~89 & 4.35~$\pm$~0.22 & $-$0.06~$\pm$~0.06 & 0.92~$\pm$~0.26  \\
61~Cyg~A       &  -- & -- & --  & -- \\
61~Cyg~B       &  -- & -- & --  & -- \\
HD~209458      &  6154~$\pm$~37 & 4.39~$\pm$~0.09 & 0.04~$\pm$~0.03  & 0.99~$\pm$~0.06 \\
$\xi$~Peg      & 6143~$\pm$~33 & 3.83~$\pm$~0.07 & $-$0.29~$\pm$~0.02 & 1.31~$\pm$~0.04 \\
HD~220009      & 4333~$\pm$~35 & 1.69~$\pm$~0.15 & $-$0.75~$\pm$~0.03 & 1.33~$\pm$~0.04 \\
\hline
\end{tabular}
\end{table*}

\begin{table*}
\caption{List of \ion{Fe}{i} and \ion{Fe}{ii} lines and spectral ranges employed in this work. The full version of this table is available at the CDS.}
\label{linelist}
\centering
\begin{tabular}{cccccc}
\hline\hline\noalign{\smallskip}
\multicolumn{2}{c}{Range} & \multicolumn{4}{c}{Atomic line}  \\
${\lambda}_{\rm min}$ & ${\lambda}_{\rm max}$ & ${\lambda}_{\rm line}$ & Species & $\chi_{\rm l}$ & $\log{gf}$ \\
{[\AA]} & [\AA] & [\AA] &  & [eV] & [dex] \\
\noalign{\smallskip}\hline\noalign{\smallskip}
\hline
\noalign{\smallskip}
4806.65 & 4811.44 & 4808.148 & \ion{Fe}{i}  & 3.252 & $-$2.690  \\ 
        &         & 4809.938 & \ion{Fe}{i}  & 3.573 & $-$2.620  \\ 
4867.96 & 4870.96 & 4869.463 & \ion{Fe}{i}  & 3.547 & $-$2.420  \\ 
4874.38 & 4879.10 & 4875.877 & \ion{Fe}{i}  & 3.332 & $-$1.900  \\ 
        &         & 4877.604 & \ion{Fe}{i}  & 2.998 & $-$3.050  \\
4880.64 & 4883.64 & 4882.143 & \ion{Fe}{i}  & 3.417 & $-$1.480  \\  
4891.36 & 4894.36 & 4892.859 & \ion{Fe}{i}  & 4.218 & $-$1.290  \\ 
4901.81 & 4906.63 & 4903.310 & \ion{Fe}{i}  & 2.882 & $-$0.903  \\ 
\ldots  & \ldots  & \ldots   & \ldots       & \ldots  & \ldots           \\    
\hline
\end{tabular}
\end{table*}
\end{appendix}

\end{document}